\PassOptionsToPackage{dvipsnames}{xcolor}
\documentclass[acmsmall]{acmart}
\AtBeginDocument{%
  }

\makeatletter   % customized
\def\@ACM@copyright@check@cc{} % customized
\makeatother % customized
\setcopyright{cc}
\setcctype{by}
\acmJournal{PACMHCI}
\acmYear{2025} \acmVolume{9} \acmNumber{7} \acmArticle{CSCW428}
\acmMonth{11}
\acmDOI{10.1145/3757609}

\usepackage{colortbl}
\usepackage{subcaption}
\usepackage{multirow}
\usepackage{makecell}
\usepackage{graphicx}
\usepackage{lscape}
\usepackage{soul}
\usepackage{enumitem}
\usepackage{tabularx}

\newcommand{\system}{\textit{HateBuffer}}

\newcommand{\uline}[1]{\setul{3pt}{0.5pt}\ul{#1}}

\definecolor{lightgray}{rgb}{0.8,0.8,0.8}
\newcommand{\roundedbox}[1]{%
  \colorbox{lightgray}{%
    \small#1%
  }%
}

\newcommand{\targetanony}{\texttt{target\_\allowbreak anonymization}}
\newcommand{\offensiveparaphrasing}{\texttt{paraphrasing\_\allowbreak offensive}}
\newcommand{\targetrevealing}{\texttt{revealing\_\allowbreak target}}
\newcommand{\offensiverevealing}{\texttt{revealing\_\allowbreak original}}

\newcommand{\fatigue}{MFSI}

\begin{document}

\title[\system{}]{\system{}: Safeguarding Content Moderators’ Mental Well-Being through Hate Speech Content Modification}
% Safeguarding Moderators’ Mental Well-being through Textual Style Modification in Hate Speech Moderation 

\author{Subin Park}
\authornote{Equal contributions.}
\email{subin.park@kaist.ac.kr}
\affiliation{%
  \institution{School of Electrical Engineering, KAIST}
  \city{Daejeon}
  \country{Republic of Korea}
}

\author{Jeonghyun Kim}
\authornotemark[1]
\email{jeonghyun.kim@kaist.ac.kr}
\affiliation{%
  \institution{School of Computing, KAIST}
  \city{Daejeon}
  \country{Republic of Korea}
}

\author{Jeanne Choi}
\email{jeanne.choi@kaist.ac.kr}
\affiliation{%
  \institution{School of Computing, KAIST}
  \city{Daejeon}
  \country{Republic of Korea}
}

\author{Joseph Seering}
\email{seering@kaist.ac.kr}
\affiliation{%
  \institution{School of Computing, KAIST}
  \city{Daejeon}
  \country{Republic of Korea}
}

\author{Uichin Lee}
\email{uclee@kaist.edu}
\affiliation{%
  \institution{School of Computing, KAIST}
  \city{Daejeon}
  \country{Republic of Korea}
}

\author{Sung-Ju Lee}
\authornote{Corresponding author.}
\email{profsj@kaist.ac.kr}
\affiliation{%
  \institution{School of Electrical Engineering, KAIST}
  \city{Daejeon}
  \country{Republic of Korea}
}

\renewcommand{\shortauthors}{Park et al.}


\begin{abstract}
Hate speech remains a persistent and unresolved challenge in online platforms. Content moderators, working on the front lines to review user-generated content and shield viewers from hate speech, often find themselves unprotected from the mental burden as they continuously engage with offensive language. 
To safeguard moderators' mental well-being, we designed \system{}, which anonymizes targets of hate speech, paraphrases offensive expressions into less offensive forms, and shows the original expressions when moderators opt to see them. 
Our user study with 80~participants consisted of a simulated hate speech moderation task set on a fictional news platform, followed by semi-structured interviews.
Although participants rated the hate severity of comments lower while using \system{}, contrary to our expectations, they did not experience improved emotion or reduced fatigue compared with the control group. In interviews, however, participants described \system{} as an effective \textit{buffer} against emotional contagion and the normalization of biased opinions in hate speech. Notably, \system{} did not compromise moderation accuracy and even contributed to a slight increase in recall. 
We explore possible explanations for the discrepancy between the perceived benefits of \system{} and its measured impact on mental well-being. We also underscore the promise of text-based content modification techniques as tools for a healthier content moderation environment. 
\end{abstract}



\received{October 2024}
\received[revised]{April 2024}
\received[accepted]{August 2024}

\maketitle

\fcolorbox{black}{white}{
  \begin{minipage}{0.98\textwidth}
    \textbf{CONTENT WARNING}: This paper contains hate speech examples, including pejorative terms, that readers may find disturbing.     
  \end{minipage}
}

% \raggedbottom % For the space between figure and body

\section{Introduction}
\label{sec:introduction}
As the volume of user-generated content continues to grow on social media~\cite{influencermarketinghub2024}, the prevalence of harmful content, particularly hate speech, has become a significant concern~\cite{unesco2023survey}. Hate speech is defined as any form of communication in speech or writing that attacks or discriminates against a person or group based on their identity (e.g.,~religion, ethnicity, nationality, race, gender, or other factors)~\cite{un2021report}, and is a long-established issue. %Per recent transparency reports from X and Facebook, 
For example, users on X reported 66.9M instances of hate speech~\cite{x_transparency_2024}, and Facebook took action on 14.6M instances of hate speech in the first half of 2024~\cite{facebook_transparency_2024}.
% According to a recent UNESCO report on public opinion~\cite{unesco2023survey}, 67 percent of 8,000 Internet users have encountered hate speech in 2023.
% reference links for recent statistics: 1) https://www.statista.com/chart/33299/online-hate-speech-encounters/ 2) https://www.statista.com/statistics/1013804/facebook-hate-speech-content-deletion-quarter/ 3) https://www.justice.gov/hatecrimes/hate-crime-statistics

% Prevalence and psychological, social impact of hate speech
Hate speech can elicit strong negative emotions~\cite{heung2024vulnerable}, resulting in serious psychological effects (e.g.,~trauma~\cite{scott2023trauma,thomas2022s} or depression~\cite{wachs2022online,tynes2008racial}) for readers, particularly the target and onlookers who identify with the targets. Given the significant harm that hate speech can cause to individuals and society at large by reinforcing harmful stereotypes~\cite{buerger2021speech,castano2021internet}, many platforms have employed commercial content moderators to review and moderate them~\cite{roberts2016commercial}. Many efforts have been developed for automated hate speech moderation, mostly based on heuristic rules~\cite{facebokblockword, instahidecomment, twitchautomod} and artificial intelligence~(AI)~\cite{micazure, googlejigsaw}. However, these tools often fail to grasp the context and subtle nuances of hate speech~\cite{alkhamissi2023token,masud2021hate,warner2012detecting,breitfeller-etal-2019-finding}. In fact, X's 2024 transparency report indicates that of the over 2M removed hate speech posts, 99.75\% required human moderators to address nuanced content, with only 0.25\% managed solely by automated systems~\cite{x_transparency_2024}. Given the persistent need for human judgment in hate speech moderation, platforms continue to recruit human moderators~\cite{indeed, ziprecruiter, amazonjobs}.
% X's content moderation 
    % To enforce our Rules, we use a combination of machine learning and human review. These systems either take action automatically, or surface content to human moderators based on user reports and/or proactive detection methods. Our human moderators use important context to make decisions about potential violations. This work is led by an international, cross-functional team with 24-hour coverage and the ability to cover multiple languages. We also have an appeals process for any potential errors that may occur.

% Content moderation is essential for combating the spread of hate speech and for maintaining safe and inclusive online spaces.
% Because of huge volumes of hate speech, many companies have implemented AI hate speech detection with different levels of hate.
% Gordon et al. found that using crowdsourced data for model evaluation can significantly overestimate models' capabilities~\cite{gordon2021disagreement}. Also, moderation models face technical challenges such as bias in classification algorithms~\cite{davidson2017automated} and the difficulty of interpreting subtle language nuances~\cite{alkhatib2019street}.
% Consequently, human moderators are still essential for manually reviewing and addressing hate speech, ensuring that moderation decisions reflect a deeper comprehension of context, tone, and cultural sensitivities~\cite{thomas2022s}.
 
These content moderators are at risk for a variety of detrimental impacts to their mental well-being due to regular exposure to harsh and offensive content, creating a challenging and emotionally demanding work environment~\cite{steiger2021psychological,roberts2019behind}. 
% Prior studies have sought to protect moderators' mental well-being by promoting workplace wellness (e.g.,~providing sabbatical or group therapy sessions)~\cite{cook2022awe, steiger2021psychological} and providing stress-relieving content (e.g.,~image of scenic landscapes, video of baby animals) during their rest~\cite{cook2022awe,lee2024alleviatevideo}. 
Recent studies have shown that modifying the style of image- and video-based content that moderators review using approaches such as grayscaling, blurring, and adding cartooning~\cite{das2020fastaccurate,karunakaran2019testing,lee2024alleviatevideo} can reduce the emotional burden of moderating harmful content without compromising accuracy. 
% safeguarding moderators' mental wellbeing while moderating hate speech 위해서 이런 content modification이 text에 적용되는것도 도움 될 수있다. however, how conent modifying approaches can adopt to text 아직 모른다. specifically, hate speech는 text기반이기 때문에 the impact stems from semantic meaning rather than visual stimuli.
However, the potential for similar modifications to text content remains underexplored. To safeguard the mental well-being of moderators who review hate speech, adapting content modification techniques for text content could be highly beneficial; however, unlike visual content, where immediate sensory stimuli often drive emotional impact, text-based hate speech derives its detrimental effect from the semantic meaning, making it challenging to directly apply visual content modification techniques. To address this, we designed \system{}, a text content modification system for hate speech moderation to alleviate moderators' mental burdens while preserving moderation performance. 

\system{} consists of four features to modify the hate speech content. First, \texttt{target\_anonymization} anonymizes the target group of potential hate speech to reduce the negative emotional impact on moderators caused by feeling attacked~\cite{cowan1996judgements} or experiencing vicarious trauma~\cite{tim2019use}. Second, \offensiveparaphrasing{} paraphrases offensive expressions to less offensive versions to prevent the emotional contagion from reviewing offensive language~\cite{cheshin2011anger, hancock2008imsad, ferrara2015measuring}. Lastly, \targetrevealing{} and \offensiverevealing{} allow moderators to optionally view the original target and offensive expressions by clicking, providing control over whether to access the full content. 

To investigate whether \system{} can reduce moderators' emotional burden without harming their performance, we aimed to answer the following research questions~(RQs):
%\setlist{nolistsep}
%\vspace{-\topsep}
\begin{itemize}[leftmargin=20px,noitemsep,topsep=6pt,partopsep=6pt]
    \item \textbf{RQ1}: How does each feature of \system{} contribute to moderators’ \textbf{mental well-being} during hate speech moderation?
    \item \textbf{RQ2}: How does each feature of \system{} influence \textbf{moderation strategies} and contribute to moderators’ \textbf{performance} in hate speech moderation?
\end{itemize}

To address these RQs, we conducted a between-subjects study with 80 participants. We distributed them into four groups: the control group as the baseline, the anonymizing group using \targetanony{}, the paraphrasing group using \offensiveparaphrasing{}, and the revealing group using \system{} with all features. For the user study, we selected 100 comments from the K-HATERS dataset~\cite{park2023k} and applied text content modification, utilizing a Large Language Model~(LLM) for \offensiveparaphrasing{}. %We conducted an experiment in which 
Participants were assigned the role of moderators to perform simulated hate speech moderation for a fictional news platform. Through surveys and semi-structured interviews, we investigated \system{}'s impact on participants' mental well-being and moderation performance. 

In contrast to our expectations, we did not detect a difference in post-study fatigue or negative emotion levels between the study conditions. However, participants in the text modification conditions rated the severity of hate speech lower than those in the control condition, and participants noted many positive aspects of the system in post-study interviews; they perceived \system{} as a buffer, providing time to prepare themselves to face the hateful content. Participants also noted that \system{} helped to protect them from normalizing biased and hateful opinions from the comments.
% perceiving hate speech comments as less hateful in the experimental groups compared with the control group.} 
In addition, despite \system{} modifying the comments by anonymizing targets and paraphrasing offensive expressions, the moderation accuracy remained similar at between 0.75--0.80 for all groups. Notably, the paraphrasing and revealing groups showed slightly higher moderation recall. 

Building on this finding, we explore possible explanations for the discrepancy between perceived benefits and the actual impact on mental well-being. Additionally, we highlight how text content modification can provide positive friction to enable a more thoughtful moderation process, and we provide considerations for adopting text content modification in commercial settings. Finally, we discuss the importance of protecting moderators' mental well-being to support a more sustainable working environment.
\section{Related Work}
\label{sec:related_work}
% 다 쓰고 리뷰할 때 생각해 봐야할 포인트: 읽다가 "그럼 이런 연구 없었나?" 라는 생각이 들지 않는지 

We review current practices in hate speech moderation, highlighting the complexity that necessitates the involvement of human moderators. We discuss the challenges human moderators face and existing approaches that address these challenges, focusing on mental well-being.

% To start with introducing the main task of our system, we investigate the definition and current strategies of hate speech moderation (Section 2.1). This section includes understanding the type of text content that we are handling in our research, hate speech, by going over the impact of hate speech on online community users and their complex nature which requires the need of human moderators. Then, we cover the challenges that human moderators face, focusing on their mental well-being and the approaches to address the challenges (Section 2.2).

\subsection{Hate Speech Moderation}
\label{sec:related:hate}
% Definition and characteristics of hate speech compared to other online harm
Hate speech is a widespread issue in online communication~\cite{das2020hate,punyajoy2023on,unesco2023survey,frances2024crack} and remains a persistent concern within HCI and CSCW. Although the exact definition of hate speech varies among countries and communities~\cite{jiang2021understanding}, it is generally defined as any form of communication, in speech or text, that attacks or discriminates against individuals or groups based on aspects of identity, such as religion, ethnicity, nationality, race, gender, or similar factors~\cite{un2021report,lupu2023offline}. 

% Prevalence and psychological, social impact of hate speech
Hate speech results in significant psychological and social harm to users within online communities~\cite{saha2019prevalance, siegel2020online}. A large body of research has reported that hate speech targeting individuals who create content on platforms, such as YouTubers, live-streamers, or journalists, can lead to considerable emotional distress~\cite{heung2024vulnerable,thomas2022s,goyal2022you, samermit2023millions, johnson2018inclusion, han2023hateraids}. This emotional harm can lead to long-term psychological consequences such as depression~\cite{wachs2022online, tynes2008racial} and trauma~\cite{scott2023trauma,thomas2022s}. The negative impact of hate speech extends beyond direct victims, affecting viewers who share the targeted identity~\cite{cowan1996judgements}. Furthermore, prevalent hate speech could foster hostility and reinforce harmful stereotypes, carrying the potential to incite violence in offline settings~\cite{buerger2021speech,castano2021internet}.
% A higher level of stress was shown from college students who were exposed to hate speech in college subreddits~\cite{saha2019prevalance}. 

% Hate speech detection and detection (linguistic features of hate speech, challenges of detection)
To address these negative impacts of hate speech, various efforts %in both industry and academia 
have been directed toward effective content moderation~\cite{gongane2022detection}. Content moderation is an organized practice of screening user-generated content posted to Internet sites, social media, and other online outlets~\cite{roberts2019behind}. Moderation is usually shaped by community guidelines and policies~\cite{redditcontentpolicy,x_community}, and content moderation may include manual review by human moderators (sometimes volunteer users~\cite{joseph2019modeartor} but often contractors hired by platforms~\cite{roberts2019behind}), semi-automated review where automated tools assist moderators, and/or fully automated methods usually based on machine learning algorithms~\cite{gongane2022detection} or hashes~\cite{farid2021overview}.
% such as the content policy of Reddit that states Reddit is ~\textit{'free of harassment, bullying, and threats of violence'}

Automated approaches have been extensively developed to facilitate large-scale content moderation. Word filtering, a traditional automated moderation technique that detects specific words or similar textual patterns violating community guidelines, is widely employed across various platforms~(e.g.,~Twitch~\cite{twitchautomod}, Facebook~\cite{facebokblockword}, Instagram~\cite{instahidecomment}, etc.). However, word filtering often falls short due to high false positive and false negative rates, stemming from its inability to interpret context~\cite{jhaver2022filterbuddy, jhaver2019human, kuo2023unsung}. In response, %researchers and engineers have been actively 
advanced AI-based moderation systems incorporate contextual understanding to improve accuracy and reduce errors~\cite{gorwa2020algorithmic,badjatiya2017deep, schmidt2017survey}.

% 이 문단에 reappropriation 과 counterspeech 이야기 해야함
Although AI-based moderation models claim high accuracy, their real-world performance often falls short. Gordon et al. noted that evaluating moderation models using crowdsourced data can dramatically overstate their capabilities~\cite{gordon2021disagreement}. For example, by adjusting for intra-annotator consistency in the popular Jigsaw toxicity task, where the model initially achieved a reported ROC AUC of 0.95, they found the performance dropped to an ROC AUC of 0.73. The complex nature of hate speech adds further challenges to detection. As hate speech targets a specific individual or group, understanding related context or background is often needed to judge whether certain content is hate speech~\cite{alkhamissi2023token, masud2021hate, alkhatib2019street}. 

% Beyond accuracy concerns, commercial moderation models frequently struggle due to their limited ability to take context into account, particularly in cases with identity-based languages, such as those based on race and gender.
Commercial moderation models frequently struggle with limited contextual understanding, particularly for identity-based language, as in cases based on race and gender~\cite{hartmann2025lost,narayanan2024ai}. This can lead to over-moderation, inadvertently flagging non-hateful content, such as counter-speech, which employs similar linguistic features to challenge or subvert discriminatory narratives~\cite{mun2024counterspeackers}. Prior research has documented various classification biases~\cite{davidson2017automated}, which further exacerbate these issues. Such biases complicate classifier adjustments, as reducing over- or under-moderation often worsens the other, particularly when moderation outcomes are unevenly distributed across identity groups~\cite{narayanan2024ai}.
Additionally, while LLMs show some potential for integration into content moderation processes, current efforts fall short in terms of accuracy and consistency~\cite{li2024hot, kolla2024llmmod}, showcasing the continued need for human moderators.

% INITIAL SUBMISSION VERSION
% Although AI-based moderation models claim high accuracy, their real-world performance often falls short. Gordon et al. noted that evaluating moderation models using crowdsourced data can dramatically overstate their capabilities~\cite{gordon2021disagreement}. For example, by adjusting for intra-annotator consistency in the popular Jigsaw toxicity task, where the model initially achieved a reported ROC AUC of 0.95, they found the performance dropped to an ROC AUC of 0.73. Moreover, the complex nature of hate speech adds further challenges to detection. As hate speech targets a specific individual or group, understanding related context or background is often needed to judge whether certain content is hate speech~\cite{alkhamissi2023token, masud2021hate, alkhatib2019street}. The limitations of the detection technology complicate the problem, with prior work documenting various types of potential bias in classification algorithms~\cite{davidson2017automated}. Moreover, while LLMs show some potential for integration into content moderation processes, current efforts fall short in terms of accuracy and consistency~\cite{li2024hot, kolla2024llmmod}, showcasing the continued need for human moderators.

% 이 문단에 TrustLab 예시 넣기. 
These limitations have led platforms such as YouTube and Spotify to take various approaches such as classifying comments into different categories~--~e.g., public, held for review, and likely spam~--~delegating final decisions on ambiguous content to human moderators~\cite{youtube2024comment,techcrunch2024spotify}. Similarly, X’s 2024 transparency report revealed that among the posts flagged for hate speech, only 0.25\% were handled automatically, with the vast majority requiring human moderators~\cite{x_transparency_2024}. TrustLab's ModerateAI follows a similar approach, using AI to pre-process content and flag potential issues, while human moderators verify automated decisions to ensure policy alignment~\cite{trustlab}. This underscores the limitations of automatic moderation and the ongoing essential role of human judgment in accurately identifying and moderating hate speech in online communities~\cite{alkhatib2019street}. 

\subsection{Mental Burden on Human Moderators: Challenges and Mitigation Strategies}
\label{sec:related:human}

% -> 그래서 노출되는 갯수 줄이거나 efficient하게 도와주려는 연구들 있음
% mental wellbeing 직접적으로 support 필요함
% style change 하는 연구 있음.. 순서로.... 
A substantial body of research has highlighted the mental burden faced by human moderators~\cite{tim2019use,casey2019trauma,steiger2021psychological,roberts2019behind}. Regular exposure to highly offensive content, including sexism, racism, and various other abuses with varying degrees of intensity~\cite{wohn2019volunteer}, places a significant psychological burden on moderators~\cite{steiger2021psychological}. This burden can result in high levels of mental distress~\cite{das2020fastaccurate,spence2024content}, burnout~\cite{spence2023psychological}, anxiety~\cite{tim2019use}, and even post-traumatic stress disorder~(PTSD)~\cite{martin2020downsides,benjelloun2020psychological,barrett2020who,gillespie2018custodians,wohn2019volunteer}. This mental strain can also drive moderators to leave their positions due to the cumulative impact of ongoing exposure~\cite{angela2024why}.

% Supports in terms of effectiveness
Various tools and services have been developed to enhance scalable, efficient, and effective moderation processes, including technological support and improvements to the work environment to alleviate the mental burden of moderators~\cite{steiger2021psychological}.
% A fundamental and commonly used approach involves word filters, enabling moderators to exclude content containing blacklisted keywords~\cite{jhaver2022filterbuddy,jhaver2019audomoderator}. %In line with this, 
% CrossMod~\cite{chandrasekharan2019crossmod} uses a large dataset of prior moderation decisions to filter explicitly offensive expressions, allowing moderators to focus only on nuanced and ambiguous contents. 
For example, a recent approach provides visual cues that direct moderators’ attention to potentially problematic content~\cite{schluger2022proactive}. Highlighting offensive language or hate speech targets identified by machine learning models has supported faster moderation~\cite{uma2023toxvis,hee2024brinjal,calabrese2024explainability,mathew2021hatexplain}. 
Additionally, providing moderators with explanations about why the content violates the community's policies can reduce the time required for conducting the moderation task~\cite{calabrese2024explainability}. Expanding on this concept, an LLM-generated description of an implied social bias in content can enrich moderators' understanding of the underlying issues, improving moderation efficiency~\cite{zhang2023biasx}.

% Additionally, Calabrese et al. provided moderators with pre-defined explanations of why specific content might violate platform policies~\cite{calabrese2024explainability}. Expanding on this concept, BiasX employs Large Language Models~(LLMs) to generate detailed insights into the social biases implied by the content, thereby enriching moderators’ understanding of the underlying issues~\cite{zhang2023biasx}.

% supports in terms of mental wellbeing
To further address the mental burden faced by moderators, %recent research has explored 
various approaches have been taken to improve mental well-being and %. Several works have emphasized 
to promote moderators' workplace wellness. These include risk mitigation strategies (e.g.,~scheduling recess, showing relaxing images), providing clinical support, and fostering peer support connections to help manage their work's psychological impacts~\cite{newton2019trauma, steiger2021psychological}. For instance, offering mindfulness content, such as meditation videos, has shown potential benefits. Lee et al. found that providing positive videos (e.g.,~scenic landscapes, baby animals) during breaks in a car accident video annotation task reduced negative emotions among moderators~\cite{lee2024alleviatevideo}. In contrast, Cook et al. observed no positive effect from using similar stimuli (cute and relaxing images) during breaks from moderating text content related to sexism, racism, and threats, indicating that visual relief alone may not fully address moderators’ emotional stress~\cite{cook2022awe}.

% karunakaran2019testing -> grayscaling, blurring 
% das2020fastaccurate -> interactive
A complementary line of work for supporting moderation processes has explored methods for reducing harm from content exposure by proactively modifying the content itself. In image moderation, Karunakaran et al. applied grayscaling and blurring to images that moderators reviewed to reduce visual stimuli, finding that this adjustment led to more positive emotional responses without compromising moderation performance~\cite{karunakaran2019testing}. Building on this, Das et al. introduced an interactive blurring intervention that allowed moderators to selectively reveal content as needed, which further reduced emotional distress for moderators~\cite{das2020fastaccurate}. Recently, similar techniques have been extended to video content moderation, incorporating blurring, grayscaling, and cartoonizing effects~\cite{lee2024alleviatevideo}. While blurring and grayscaling maintained comparable moderation accuracy to the baseline, cartoonizing was perceived as an effective intervention for reducing negative emotions when dealing with provocative and unpleasant videos.
% cartooning enhanced moderators' emotional protection.
% If this intervention technique were implemented, would it help reduce your negative emotions when dealing with provocative or unpleasant vidoes? -> highest proportion of storngly agree. 
% cartoonizing was the most impressive (What intervention technique impressed you the most? Additionally, what were the reasons for yout choice? )

Inspired by the existing work on modifying the content to alleviate moderators’ mental burdens, we explore how to realize that for user-generated textual content. In this work, we propose \system{}, a text-based content modification system for hate speech moderation. While previous research has focused on limiting moderators’ exposure to visual stimuli by altering images or videos, our work focuses on textual content by introducing novel content modification techniques. Given the essential role that human moderators play in online spaces and the unique challenges they encounter, we explore the impact of \system{}'s features on their mental well-being and moderation effectiveness.
\section{\system{}}
\label{sec:method}

% \subsection{Design of \system{}'s Features}
\begin{figure}[!t]
    \centering
    \begin{minipage}[b]{0.5\linewidth}
        \centering
        \begin{subfigure}[b]{\linewidth}
            \centering
            \includegraphics[width=\textwidth]{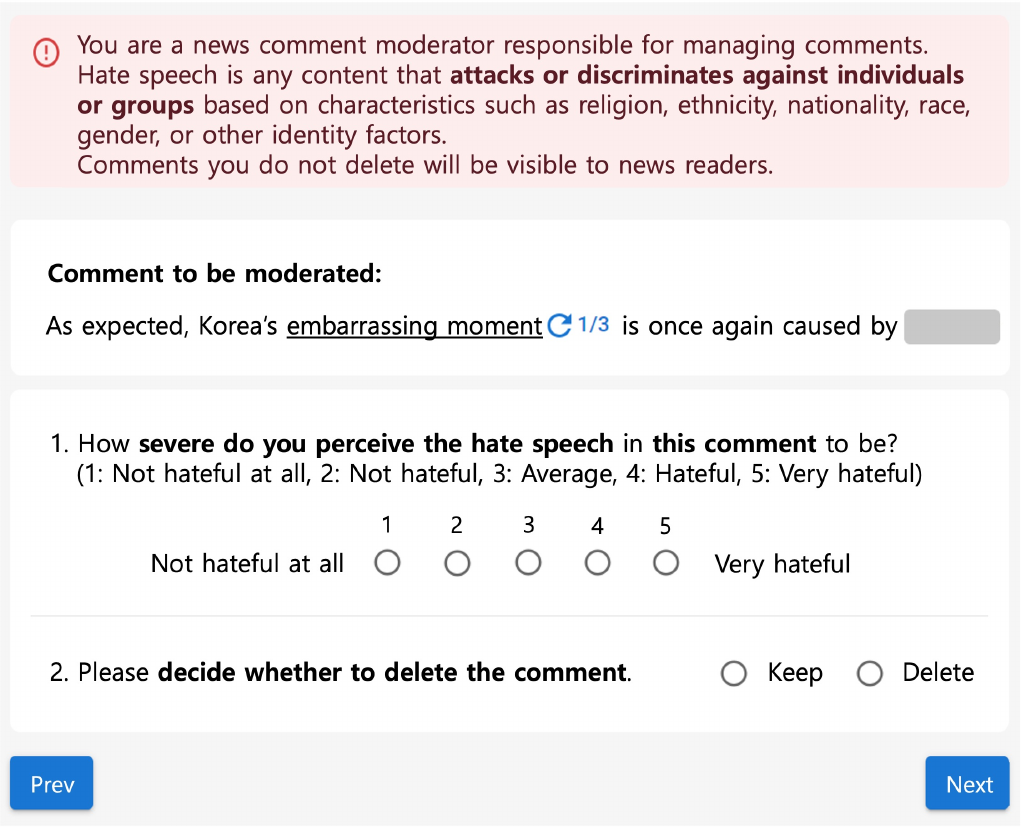}
            \vspace{-15pt}
            \caption{Default view.}
            \label{fig:design:default}
        \end{subfigure}
    \end{minipage}%
    \hfill
    \begin{minipage}[b]{0.48\linewidth}
        \vfill
        \centering
        \vspace{0.1\linewidth}
        \begin{subfigure}[b]{\linewidth}
            \centering
            \includegraphics[width=\textwidth]{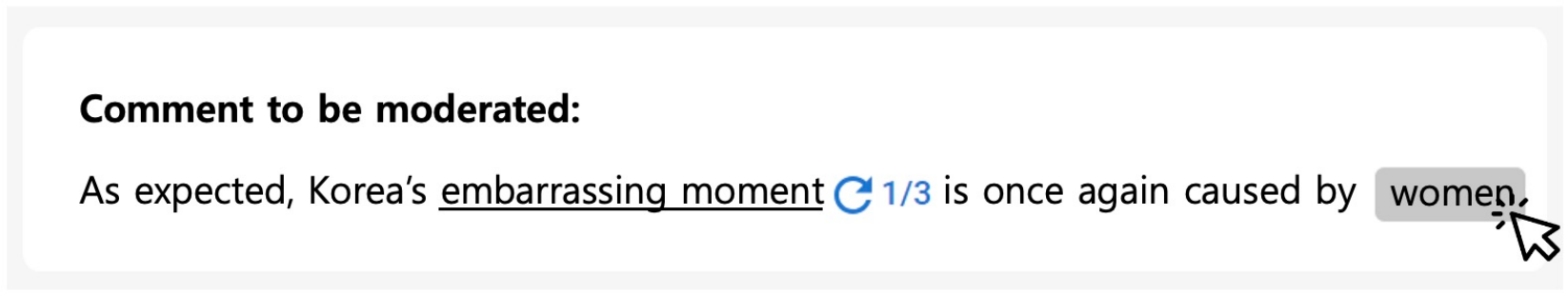}
            \vspace{-15pt}
            \caption{\targetrevealing{}.}
            \label{fig:design:target}
        \end{subfigure}
        \vfill \vspace{.15\linewidth}
        \begin{subfigure}[b]{\linewidth}
            \centering
            \includegraphics[width=\textwidth]{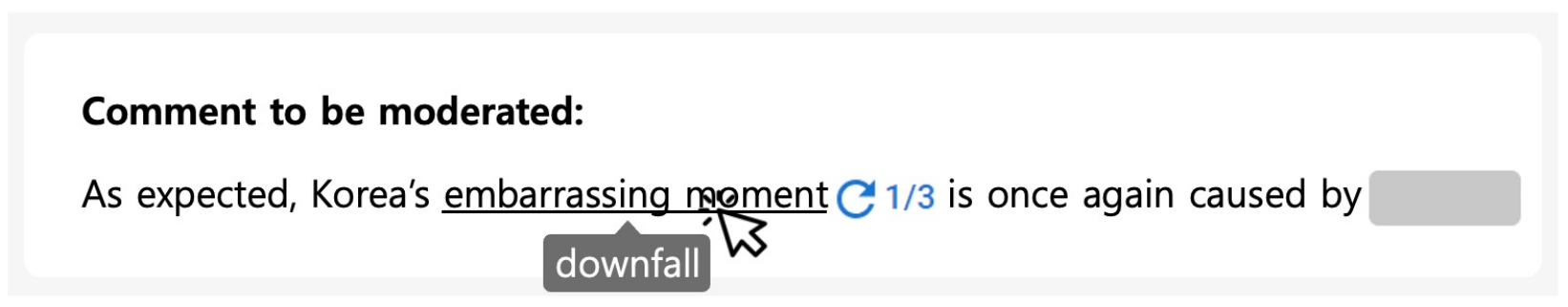}
            \vspace{-15pt}
            \caption{\offensiverevealing{}.}
            \label{fig:design:offensive}
        \end{subfigure}
        \vspace{0.12\linewidth}
    \end{minipage}
    \vspace{-5pt}
    \caption{Screenshots of \system{}. Instructions for the moderation task, a comment to moderate, and an interface for moderation are given. (a)~By default, \system{} provides a modified comment with \targetanony{} and \offensiveparaphrasing{}. (b)~By clicking the anonymized target, moderators can see the original target expression (\targetrevealing{}). (c)~By clicking the paraphrased expression, moderators can see the original offensive expression (\offensiverevealing{}).}
    \label{fig:design}
    % \vspace*{-0.4cm}
\end{figure}

We propose \system{}, a system designed to support moderators’ mental well-being during hate speech moderation~(Fig.~\ref{fig:design}). \system{} consists of four features: \targetanony{}, \offensiveparaphrasing{}, \targetrevealing{}, and \offensiverevealing{}. We present each feature and its design rationale. 
% To safeguard moderators' mental well-being, we designed \system{}, which anonymizes targets of hate speech, paraphrases offensive expressions into less offensive forms, and shows the original expressions when moderators opt to see them. 

% Given that visually changing the text content (e.g.,~grayscaling~\cite{}, cartoonizing~\cite{}) does not change the amount of semantic delivered to moderators, 
% Inspired by previous work that reduces moderators’ exposure to harmful visual content by changing the style (e.g.,~grayscaling~\cite{das2020fastaccurate,karunakaran2019testing} and cartoonizing~\cite{lee2024alleviatevideo}), we designed two linguistic style change features: \targetanony{} and \offensiveparaphrasing{}. Furthermore, building on the concept of positive friction, we introduced two interactive features: \targetrevealing{} and \offensiverevealing{}. 

\subsection*{Target Anonymization} Moderators frequently encounter hate speech that targets their identities or communities, which can have a negative emotional impact~\cite{cowan1996judgements}. Even when the hate speech is not personally directed at them, offensive language aimed at their social group or identity can still evoke secondary trauma or PTSD~\cite{steiger2021psychological}. To mitigate the emotional impact of moderating hate speech, we designed \targetanony{}. As shown in Fig.~\ref{fig:design:default}, it anonymizes the original target expression `women' using a gray cover. 
% As shown in Figure~\ref{fig:design:target}, this intervention anonymizes the target of hate speech with a gray placeholder. 

\subsection*{Paraphrasing Offensive Expressions} %The emotional impact of language has been widely observed~\cite{bargh1996automaticity, guillory2011upset, herrando2021emotional, goldenberg2019digital}. Especially, r
Reading %negative expressions, such as 
offensive expressions can evoke negative sentiment through the behavioral phenomenon of emotional contagion~\cite{bargh1996automaticity, guillory2011upset, herrando2021emotional, goldenberg2019digital}. Research has shown that emotional contagion, usually examined through non-verbal cues such as facial expressions~\cite{herrando2021emotional}, can also occur through text~\cite{cheshin2011anger, hancock2008imsad, ferrara2015measuring}. These expressions not only cause immediate emotional responses but also tend to linger in memory, resulting in an enduring cognitive effect~\cite{thomas2022s, scott2023trauma}. To mitigate these impacts on content moderators, we designed a feature to reduce exposure to offensive language. % and safeguard mental well-being. % revise while finding references

One approach to reducing the negative impact of text involves using euphemism, which replaces direct, potentially harmful language with softer alternatives to create emotional distance and reduce the shock factor~\cite{mayor2009longman}. Euphemisms have been observed to obscure the harshness of the original content, thereby lowering the emotional intensity of the message~\cite{burkhardt2010euphemism}. Prior research has shown that, even when the content remains the same, the tone of expression can significantly alter the reader’s perception and emotional response~\cite{hu2018touch}. % add more citation regarding euphemistic expressions

With this in mind, we designed \offensiveparaphrasing{}, which paraphrases offensive expressions within hate speech into less offensive versions. As shown in Fig.~\ref{fig:design:default}, it paraphrases the original offensive expression `downfall' to a less offensive version, `embarrassing moment.'\footnote{Note that examples provided throughout this paper have been translated by the research team from their original Korean into English. We have endeavored to find English phrasings that capture the same sentiment, but this is difficult in some cases due to cultural differences. Participants in the study saw the phrases and their modifications in the original Korean.} Since we replaced only specific expressions rather than entire sentences, the paraphrased terms may occasionally feel out of place in the surrounding context, even if the overall meaning is preserved. For better readability, \system{} provides up to three alternatives shown as `1/3' next to the underlined text, to ensure a smoother integration within the comment. We used an LLM to achieve this goal, and the detailed process of generating the paraphrased expressions is described in~\S\ref{sec:user:curation}.
% To ensure the appropriateness of the paraphrased expressions, \system{} provides up to three alternative expressions, allowing flexibility in cases where the initial paraphrase may not fully represent the original sentence’s context. 
% This intervention aims to lessen the emotional burden on moderators by altering the linguistic tone while retaining the core meaning of the original content.

% better readability
% seamless align
% For better readability of modified comments, despite of replacing certain expression to paraphsed expression
% 우리가 문장 전체를 바꾼게 아니고 특정 표현만 페러프레이즈해서 넣었기 때문에, 의미가 통하더라도 주변문맥상 어색할 수 있다. for better readability, \system{} provides up to three alternative expressions

% revealing
\subsection*{Revealing Targets or Original Offensive Expressions}
Research on online harassment has shown that providing users with a high-level summary of negative responses to their posts allowed them to engage with the content that elicited negative emotions on an optional basis. Interestingly, even when users opted to read the feedback, the summary helped them mentally prepare for the negative content. Participants reported feeling less emotionally impacted when they could anticipate the criticism before fully engaging with it. This strategy highlights the importance of introducing a preparatory phase, giving users a moment to prepare before exposure to potentially harmful materials~\cite{kim2024respect}.

This approach aligns with the idea of positive friction, which refers to intentional interventions designed to momentarily interrupt automatic processes and encourage mindfulness and deliberate decision-making~\cite{chen2024exploring}. For instance, trigger warnings operate as positive friction by prompting users to decide whether they want to engage with sensitive content that might negatively affect their emotional state~\cite{haimson2020trans,yukari2018selfharm}.

Building on this idea, we developed two features: \targetrevealing{} and \offensiverevealing{}. \targetrevealing{} shows the original target of the hateful content when the moderators click the anonymized target expression~(Fig.~\ref{fig:design:target}). Similarly, \offensiverevealing{} displays the original offensive expression when the moderator clicks the paraphrased offensive expression~(Fig.~\ref{fig:design:offensive}). By offering these options, we %aim to 
create a psychological buffer that reduces moderators' emotional strain while enabling them to effectively carry out their moderation tasks. 

We implemented \system{} as a web application using TypeScript and React.js, with real-time logging on Firebase Firestore~\cite{firestore}.
% By offering moderators the autonomy to choose when to reveal sensitive content, we aim to introduce a psychological buffer that helps reduce emotional strain.

% \subsection{Implementation}
% We implemented a web-based application to conduct the user study. Front-end components were implemented using Typescript with React.js and communicated with the server to fetch experimental data, send experimental logs, and monitor the participants' progress in the experiment. We collected experimental logs in real-time using Firebase.
\section{User Study}
\label{sec:user}
In this section we detail the user study methods, including data curation, evaluation metrics, study design, and analyses used to assess \system{}'s support for moderators' well-being and performance in hate speech moderation tasks.
% In this section, we describe our user study, including the data curation process, the experimental design, and the measures we collected to evaluate the effectiveness of \system{} in supporting both the mental well-being of hate speech moderators and their moderation performance. We also detail the implementation of the experimental platform, participant recruitment, and the statistical and qualitative analyses conducted to understand participants' experiences. Ethical considerations and procedural steps are outlined to ensure compliance with institutional guidelines.

\subsection{Study Setup}
\label{sec:user:setup}

\begin{figure}[t]
\begin{minipage}{\textwidth}
\centering
    \subfloat[The control group.]{\includegraphics[width=.6\textwidth]{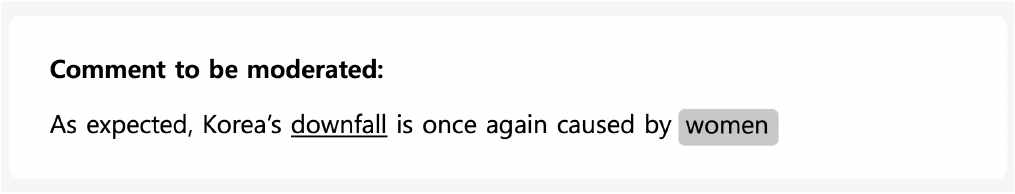}\vspace{-5pt}\label{fig:screenshot:control}}
    \vspace{5pt}
    \hfill
    \subfloat[The anonymizing group.]{\includegraphics[width=.6\textwidth]{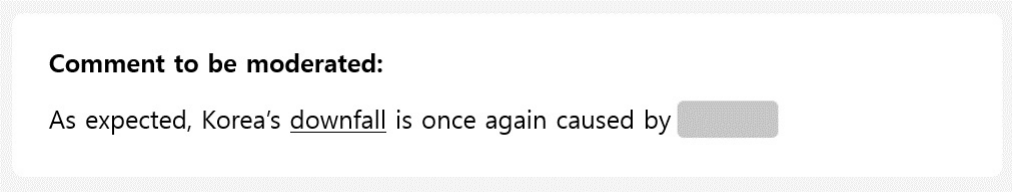}\vspace{-5pt}\label{fig:screenshot:target}}
    \vspace{5pt}
    \hfill
    \subfloat[The paraphrasing group.]{\includegraphics[width=.6\textwidth]{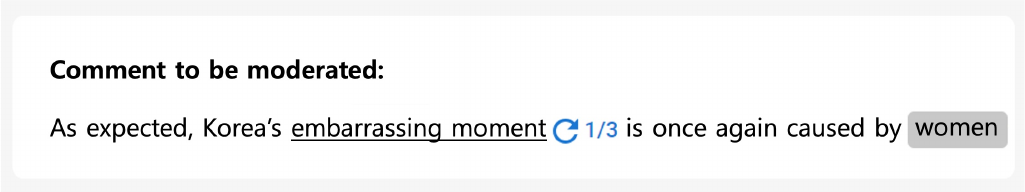}\vspace{-5pt}\label{fig:screenshot:offensive}}
    \vspace{-5pt}
    \caption{Screenshots of the user interface with an example comment for (a)~control, (b)~anonymizing, and (c)~paraphrasing groups.}
    \label{fig:group}
\end{minipage}
\end{figure}

We designed a between-subjects user study that included a hate speech moderation experiment followed by semi-structured interviews to explore how \system{} supports moderators in protecting their mental well-being and moderation task performance. To clearly observe the effectiveness of \targetanony{} and \offensiveparaphrasing{}, we designed four study groups: the control group as the baseline, the anonymizing group using only \targetanony{}, the paraphrasing group using only \offensiveparaphrasing{}, and the revealing group using \system{}, where participants initially encounter anonymized targets and softened expressions but can reveal the target and original expressions using \targetrevealing{} and \offensiverevealing{}.
%includes all four features of  \targetanony{}, \offensiveparaphrasing{}, \targetrevealing{}, and \offensiverevealing{}. 

For the control group, target expressions are highlighted in gray and offensive expressions are underlined (Fig.~\ref{fig:screenshot:control}). This design reflects the standard visual support typically employed in collaborative moderation support systems, both in real-world platforms~\cite{discordAutoModeration,sightengineTextModeration} and academic research~\cite{uma2023toxvis,hee2024brinjal,calabrese2024explainability,choi2023convex,park2016supporting,song2023modsandbox}.
% For the control group, target expressions are highlighted with a gray background and offensive expressions are underlined, similar to the baseline level of visual support commonly used in collaborative moderator support systems~\cite{uma2023toxvis,hee2024brinjal,calabrese2024explainability} (Fig.~\ref{fig:screenshot:control}). \jh{This design choice was informed by both academic literature and real-world moderation practices. For instance, volunteer moderator tools like Discord's AutoMod~\cite{discordAutoModeration} visually mark filtered phrases and show the matching rule, while commercial API such as Sightengine~\cite{sightengineTextModeration} annotate and categorize offensive content~(e.g., sexual, violent) to assist human moderators. In addition, several academic systems, such as ConvEx~\cite{choi2023convex}, CommentIQ~\cite{park2016supporting}, and ModSandbox~\cite{song2023modsandbox}, have employed similar visual cues to support moderation tasks. These precedents validate our use of gray highlighting and underlining as baseline visual support.}
For the anonymizing group, only \targetanony{} is given, where the target expression is anonymized with a gray cover (Fig.~\ref{fig:screenshot:target}). For the paraphrasing group, only \offensiveparaphrasing{} is given, where the offensive expression is paraphrased as less offensive (Fig.~\ref{fig:screenshot:offensive}). Lastly, for the revealing group, the full set of features in \system{} is given.
% In the following sections, we describe the data curation process for the simulated hate speech moderation, the measures we collected, the overall study procedure, the implementation of the experiment webpage, participant details, the analyses conducted, and ethical considerations.

\subsection{Data Curation}
\label{sec:user:curation}

% We curated 100 comments for moderation experiment by utilizing the K-HATERS dataset~\cite{park2023k}, the largest Korean hate speech news comment dataset with human-labeled hate class (i.e.,~hate speech or normal) with target and offensive expressions. 
To curate 100 comments for a moderation experiment, we reviewed Korean hate speech datasets: K-HATERS~\cite{park2023k}, K-MHaS~\cite{kmhas2022lee}, BEEP!~\cite{beep2022moon}, and KoLD~\cite{kold2022jeong}. From these, we selected the largest dataset that included sufficiently detailed labeling criteria and rich annotations (e.g., topic, target, and offensive rationales). K-HATERS is the largest Korean hate speech dataset with human-labeled hate classes (i.e.,~hate speech or normal) and labels for target and offensive expressions~\cite{park2023k}. The dataset defines hate speech as \textit{``words or phrases containing aggression or derogatory remarks directed at individuals or groups with specific attributes''} and categorized comments into 11 topics (e.g., insult, violence, sexual hate, and race) similar to hate speech policies used by popular social media platforms~\cite{x_community,meta_hatefulconduct,reddit_hatefulconduct}.

We began by reviewing randomly selected 300 hate speech comments and 300 normal comments with diverse topic labels (i.e.,~gender, politics, region, and job) from the entire dataset. The first and second authors individually reviewed and selected comments that provide sufficient context to be understood without reading the corresponding news article, as the dataset consists of news comments. We also verified that the hate class, target expressions, offensive expressions, and topics of the comments were correctly labeled. The final selection consisted of 50 hate speech comments and 50 normal comments, chosen based on agreements between at least two authors. Fleiss' Kappa score reached 0.88, indicating almost perfect agreement~\cite{landis1977measurement}. Additionally, to prevent participants from moderating hate speech without reading the comments using annotations of target and offensive expressions as cues, we added annotations for key subjects/objects (instead of target expressions) and keywords (instead of offensive expressions) in the normal, non-hate speech comments through iterative discussions.

To apply \offensiveparaphrasing{} intervention on curated comments, we paraphrased the offensive expressions using LLM. We used GPT-4o, chosen for its fluency in Korean and effectiveness in rephrasing~\cite{gpt4o}. We generated ten paraphrased versions of each comment by prompting the LLM as follows: (1)~assigning the role of a text content moderator for Korean news comments, (2)~explaining the definition of euphemism, (3)~requesting the paraphrasing of only the annotated offensive expressions euphemistically, and (4)~ensuring the original meaning remained unchanged (see Supplementary material~\ref{sec:supp:prompt} for the full prompt). We then filtered the paraphrased comments, retaining only those with a similarity score higher than 0.7 using cosine similarity of two text embeddings~(OpenAI text-embedding-3-small~\cite{embedding}) by referring to the criteria used by LLM-based text data augmentation works~\cite{jayawardena2024parafusion,tripto2023ship}. Afterward, the two lead authors selected three paraphrased comments for each original comment that best preserved the original meaning while still providing alternative expressions. We followed the same process for normal comments but paraphrased them into similar expressions.   

% results
In addition to the 100 comments to be used in the experiment, we selected eight more hate speech comments to be used to obtain the hate sensitivity level of each participant as part of the recruitment process, following the same process we used for the main data curation but without the paraphrasing process. 

\subsection{Measures}
\label{sec:user:measures}
To investigate the effectiveness of \system{} in protecting hate speech moderators' mental well-being, we collected quantitative measures across two dimensions: mental well-being and moderation performance.  

\subsubsection{Mental Well-being}
To understand how \system{} supports participants in preserving their mental well-being during hate speech moderation, we collected perceived hate severity, perceived effectiveness in mental well-being protection, Scale of Positive and Negative Experience~(SPANE) score, and Multidimensional Fatigue Symptom Inventory~(\fatigue{}). The detailed explanations of each measure are as follows: 

\begin{itemize}[leftmargin=20px,noitemsep,topsep=4pt,partopsep=4pt]
    \item \textbf{Perceived Hate Severity.} To understand whether \system{} reduces the offensiveness of comments, we observed how participants perceived the hate severity of the comments they saw. We asked them to evaluate the hate severity of each comment during the main task: ``How severe do you perceive the hate speech in this comment to be?'' in a 5-point Likert scale (1:~Not hateful at all, 5:~Very hateful). 
    \item \textbf{Perceived Effectiveness in Mental Well-Being Protection.} We measured how participants perceived the effectiveness of \system{} in protecting their mental well-being. We asked them to rate the statement, ``\{feature\} helped protect my mental well-being,'' on a 5-point Likert scale (1:~Strongly disagree, 5:~Strongly agree) after the experiment.
    \item \textbf{SPANE.} We examined how moderating hate speech with \system{} affected participants’ emotions differently using the Scale of Positive and Negative Experience~(SPANE)~\cite{diener2010new}. Participants responded to the adjusted question, ``To what extent do you feel at this moment?'', on six positive items (e.g.,~Pleasant, Happy) and six negative items (e.g.,~Unpleasant, Sad) on a 5-point Likert scale (1:~Not at all, 2:~A little, 3:~Moderately, 4:~Quite a bit, 5:~Extremely) before and after the experiment. SPANE\_B, the balance of positive and negative, is calculated by subtracting the sum of negative item scores from the sum of positive item scores, ranging from -24 to +24, indicating that a positive value represents participants feeling more positive than negative emotions.
    \item \textbf{\fatigue{}.} Given the mentally demanding nature of hate speech moderation, we assessed the fatigue caused by moderating hate speech with \system{}. 
    %Measuring fatigue in this study is vital due to the mental demands of moderation. To assess fatigue scales after moderating hate speech with style change interventions, 
    We used the Multidimensional Fatigue Symptom Inventory-Short Form~(\fatigue{}-SF)~\cite{stein2004further}, focusing on emotional, mental, and vigor subscales before and after the experiment. We asked ``Which of the following best describes how true each statement is for you at this moment?'' on a 5-point Likert scale (1:~Not at all, 2:~A little, 3:~Moderately, 4:~Quite a bit, 5:~Extremely). \fatigue{} is calculated by subtracting the sum of emotional and mental scores from the sum of vigor scores, ranging from -28 to 54, with higher values indicating greater fatigue. 
    % This approach excludes general and physical fatigue, aligning with the cognitive nature of the task.
\end{itemize}

\subsubsection{Moderation Performance}
\label{sec:user:measure:performance}
To evaluate \system{}'s feasibility as a moderation support tool, we collected perceived effectiveness in hate speech moderation, moderation accuracy, moderation recall, and task completion time. The detailed explanation of each measure is as follows: 

\begin{itemize}[noitemsep,leftmargin=20px,topsep=4pt,partopsep=4pt]
    \item \textbf{Perceived Effectiveness in Hate Speech Moderation.} We assessed how participants perceived each part of \system{}'s effectiveness for moderating hate speech. We asked them to rate the statement, ``\{feature\} helped perform the moderation task,'' on a 5-point Likert scale (1:~Strongly disagree, 5:~Strongly agree) after the experiment.
    \item \textbf{Moderation Accuracy and Recall}. We measured moderation accuracy and recall to compare how consistently participants moderated hate speech with \system{}. These statistics are considered here more as baselines for comparison rather than objective indicators of real-world performance. During the main task, participants made moderation decisions---either `delete' or `keep'---for each comment. We then calculated moderation accuracy and recall based on their responses. Moderation accuracy, defined as the proportion of participants' decisions consistent with the labeled dataset, captures how reliably participants identified what should or should not be deleted. Moderation recall refers to the ratio of deleted hate speech comments to the total number of true hate speech comments.
    \item \textbf{Task Completion Time.} We evaluated the time efficiency of using \system{} for moderating hate speech by collecting the task completion time for moderating 100 comments.
\end{itemize}

\subsection{Participants}
\label{sec:method:participants}
\begin{table}[t]
\caption{Statistical summary of participants' demographics of each group. Moderators \# represent the number of participants who have served as a moderator on the online platform~(e.g., Facebook groups, YouTube channels, Naver Cafes, etc.).}
\label{tab:demographic}
\resizebox{\textwidth}{!}{
\begin{tabular}{ccccclccccc}
\Xhline{2\arrayrulewidth}
\multirow{2}{*}{\textbf{Group}} & \multirow{2}{*}{\textbf{Num}} & \multicolumn{3}{c}{\textbf{Age}}            &  & \multicolumn{3}{c}{\textbf{Gender}}                      & \multirow{2}{*}{\textbf{Moderators \#}} & \multirow{2}{*}{\textbf{\begin{tabular}[c]{@{}c@{}}Hate\\ Sensitivity\end{tabular}}} \\ \cline{3-5} \cline{7-9}
\multicolumn{1}{l}{} &                               & \textbf{Mean} & \textbf{Min} & \textbf{Max} &  & \textbf{Female} & \textbf{Male} & \textbf{No disclosure} &                                        &                                                                                      \\ \hline
\textbf{Control}     & 20                            & 23.05±2.35    & 18           & 27           &  & 8               & 12            & 0                      & 3                                      & 3.94±0.70                                                                            \\
\textbf{Anonymizing}      & 20                            & 24.75±7.12    & 19           & 51           &  & 7               & 12            & 1                      & 2                                      & 3.98±0.75                                                                            \\
\textbf{Paraphrasing}   & 20                            & 24.70±7.02    & 18           & 49           &  & 10              & 10            & 0                      & 2                                      & 4.08±0.71                                                                            \\
\textbf{Revealing}   & 20                            & 24.55±4.01    & 19           & 33           &  & 8               & 11            & 1                      & 4                                      & 4.05±0.72                                                                           \\ \Xhline{2\arrayrulewidth}
\end{tabular}}
\end{table}

\begin{table}[t]
\caption{Interview participants' demographic and moderator experience information. P represents participant number, G represents gender, and M represents moderator experience.}
\label{tab:interviewee}
\resizebox{.8\textwidth}{!}{
\begin{tabular}{ccccccccccccccccccc}
\Xhline{2\arrayrulewidth}
\multicolumn{4}{c}{\textbf{Control group}}                                  &  & \multicolumn{4}{c}{\textbf{Anonymizing group}}                                  &  & \multicolumn{4}{c}{\textbf{Paraphrasing group}}                                   &  & \multicolumn{4}{c}{\textbf{Revealing group}}                                  \\ \cline{1-4} \cline{6-9} \cline{11-14} \cline{16-19} 
\textbf{P}     & \textbf{Age} & \textbf{G} & \multicolumn{1}{c}{\textbf{M}} &  & \textbf{P}    & \textbf{Age} & \textbf{G} & \multicolumn{1}{c}{\textbf{M}} &  & \textbf{P}       & \textbf{Age} & \textbf{G} & \multicolumn{1}{c}{\textbf{M}} &  & \textbf{P}       & \textbf{Age} & \textbf{G} & \multicolumn{1}{c}{\textbf{M}} \\ \cline{1-4} \cline{6-9} \cline{11-14} \cline{16-19} 
C01 & 18           & F          &                                &  & A01 & 19           & F          &                                &  & P01 & 20           & F           &                                &  & R01 & 19           & M          &                                \\
C02 & 21           & F          &                                &  & A02 & 20           & M          &                                &  & P02 & 20           & F           & \multicolumn{1}{c}{\checkmark} &  & R02 & 20           & M          &                                \\
C03 & 23           & M          &                                &  & A03 & 21           & F          &                                &  & P03 & 22           & F           &                                &  & R03 & 21           & M          & \multicolumn{1}{c}{\checkmark} \\
C04 & 23           & M          &                                &  & A04 & 22           & M          &                                &  & P04 & 23           & F           &                                &  & R04 & 21           & F          &                                \\
C05 & 23           & M          & \multicolumn{1}{c}{\checkmark} &  & A05 & 23           & M          &                                &  & P05 & 23           & F           &                                &  & R05 & 24           & F          & \multicolumn{1}{c}{\checkmark} \\
C06 & 23           & F          &                                &  & A06 & 24           & M          &                                &  & P06 & 26           & M           & \multicolumn{1}{c}{\checkmark} &  & R06 & 28           & M          &                                \\
C07 & 23           & F          &                                &  & A07 & 25           & F          &                                &  & P07 & 26           & M           &                                &  & R07 & 28           & F          &                                \\
C08 & 24           & M          &                                &  & A08 & 26           & M          & \multicolumn{1}{c}{\checkmark} &  & P08 & 28           & M           &                                &  & R08 & 29           & F          & \multicolumn{1}{c}{\checkmark} \\
C09 & 24           & M          & \multicolumn{1}{c}{\checkmark} &  & A09 & 28           & M          & \multicolumn{1}{c}{\checkmark} &  & P09 & 32           & M           &                                &  & R09 & 32           & M          & \multicolumn{1}{c}{\checkmark} \\
C10 & 25           & F          & \multicolumn{1}{c}{\checkmark} &  & A10 & 30           & M          &                                &  & P10 & 49           & F           &                                &  & R10 & 33           & M          &                               \\\Xhline{2\arrayrulewidth}
\end{tabular}
}
\end{table}

We recruited 80 participants by uploading the recruitment post to our institution's online communities and sending cold emails to moderators of 139 active Naver Cafes.\footnote{Naver Cafe (\url{https://section.cafe.naver.com/ca-fe/home}) is one of the most popular online community platforms in South Korea.} To be eligible for the study, participants had to be (1)~over 18 years old and (2)~fluent in Korean. We calculated each participant’s hate sensitivity score by asking them to rate the severity of eight hate speech comments---including two comments from each of four categories: gender, politics, region, and job---and then averaging their ratings to determine their overall hate sensitivity level. We utilized hate sensitivity scores to evenly distribute participants across groups. To minimize potential biases, we balanced groups based on participants' age and gender. Despite our best efforts, minor imbalances inevitably occurred due to unexpected factors, such as last-minute scheduling changes from participants. However, we ensured these differences were minimal and did not significantly affect group comparability. To confirm this, we conducted a Kruskal-Wallis test specifically on participants' hate sensitivity scores across the groups, which revealed no statistically significant difference ($H$=2.55,~$p$=.466).

Table~\ref{tab:demographic} presents the statistical summary of participants across the four groups. The average age was 24.26 years (min$=$18;~max$=$51;~std$=$5.45). Of the participants, 33 (41.25\%) identified as female, 45 (56.25\%) as male, and two chose not to disclose. Eleven participants (13.75\%) had prior experience as content moderators for various platforms (e.g.,~Facebook groups, YouTube channels, Naver Cafes, etc.). Table~\ref{tab:interviewee} shows the detailed demographic and moderator experience of the interviewees in each group. To gather more meaningful insights from a moderator’s perspective, we included those with moderation experience as interviewees, and we randomly selected additional participants from each group to ensure that half of the group participated in the interviews. Participants were compensated with 20,000 KRW (approximately USD 14.50) for the 1-hour user study and an additional 10,000 KRW (approximately 7.25 USD) if they participated in the subsequent half-hour interview. 
% Interview particiapnts (more detail demographic table with P_num) 
    % how we selected: include moderators, balanced gender, age 
% Half of the participants were selected for interviews, with a higher priority given to those who had prior moderator experience. Additionally, we aimed to balance the sample with respect to gender, age, and levels of hate severity to ensure a diverse and representative demographic. A detailed demographic table of interviewees, including participant numbers (P\_num), is provided for further insights.

\subsection{Study Procedure}
\label{sec:user:procedure}

\begin{figure}[t]
    \centering
    \includegraphics[width=\textwidth]{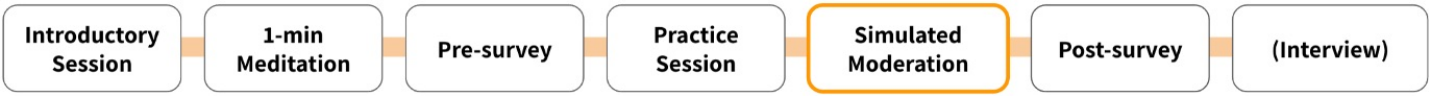} 
    \vspace{-15pt}
    \caption{Overall user study procedure. Half of the participants from each group were interviewed.}
    \label{fig:procedure}
\end{figure}

Fig.~\ref{fig:procedure} shows the overall procedure of our user study, which was consistent across the four groups. First, we delivered definitions for hate speech in the introductory session and explained content moderators' roles and tasks during the experiment. To ensure participants understood how to use the features available in their assigned condition, we presented dummy comments~(e.g.,~\textit{``In the comment, the \roundedbox{targets} are highlighted in gray, and \uline{offensive expressions} are underlined.''}) and explained each feature in detail. After the explanation, participants were given the chance to ask clarifying questions. We guided participants to follow a one-minute meditation video to allow them to begin the study with a neutral emotional state. Next, participants responded to the pre-survey with the SPANE and \fatigue{} questionnaire and practiced with group-specific features on dummy comments introduced during the introductory session.
% interacted with the intervention of the group they allocated on dummy comments (e.g.,~``In the comment, the targets are highlighted in gray, and offensive expressions are underlined.'') as a practice. 

% In a simulated moderation experiment, participants were asked to evaluate the perceived hate severity and moderate hate speech comments among 100 comments. Participants were assigned the role of moderators to perform simulated hate speech moderation for a fictional news platform. 
In a simulated moderation experiment for a fictional news platform, participants were assigned as moderators and asked to assess the perceived hate severity and moderate hate speech within a set of 100 comments.
After completing the task, participants completed a post-survey, which asked SPANE, \fatigue{}, and questions about the perceived effectiveness of \system{} in protecting mental well-being and moderating hate speech. Finally, we selectively interviewed participants to understand their experiences and perceptions of the content modification features. We interviewed participants, focusing on their perceptions of \system{} and each feature in supporting mental well-being and moderation performance, as well as their experiences during the simulated hate speech moderation experiment. The two lead authors conducted the user study online via Zoom. Up to four participants participated in each 1-hour session, with a maximum of two participants pre-selected during the scheduling phase for an additional half-hour one-on-one interview. This arrangement allowed two interviews to be conducted simultaneously using Zoom’s breakout room feature. Since the study did not require face-to-face interaction, participants were allowed to turn off their cameras. The direct message function of the chat was used for participants to ask any help if needed during the study. Participants did not interact with each other at any point. We share the survey questionnaire and interview protocol in Supplementary material~\ref{sec:supp:survey} and~\ref{sec:supp:interview}, respectively.

\subsection{Analysis}
\label{sec:method:analysis}
% Our analysis aimed to identify patterns and correlations between the moderators' psychological states and their engagement with hate speech content. 
We used a mixed-methods approach, combining qualitative and quantitative data to comprehensively understand the participants' experience with content modification interventions.

\subsubsection{Quantitative Analysis}
We conducted a descriptive statistical analysis on the collected measures: perceived hate severity, SPANE\_B, \fatigue{}, moderation accuracy and recall, and task completion time. To account for personal bias in perceived hate severity, we applied z-score normalization. We confirmed normality using the Shapiro-Wilk test. If the data followed a normal distribution, we conducted a one-way ANOVA; otherwise,  a Kruskal-Wallis test, to observe significant differences across groups. We performed two-tailed t-tests or Mann-Whitney U tests for pairwise comparisons. All between-subject analysis p-values were corrected with Bonferroni correction to control for multiple comparisons. We conducted Wilcoxon signed-rank tests for within-subject comparisons of SPANE\_B and \fatigue{} changes within each group.

% For the quantitative results, we used statistical methods to analyze the survey responses and experimental logs. Descriptive statistics provided an overview of the participants' mental well-being, moderation performance, and efforts to moderate hate speech with style change interventions. We assessed the normality of perceived hate severity, SPANE questionnaire scores, moderation accuracy and sensitivity, task completion time, and  \fatigue{} questionnaire scores using the Shapiro-Wilk test. Based on these results, we analyzed whether there were differences between groups using either the one-way ANOVA test or the Kruskal-Wallis test.

% Qualitative analysis
\subsubsection{Qualitative Analysis}
We conducted inductive thematic analysis~\cite{braun2006using} on the participants' interview data to understand participants' perceptions of interventions and moderation experiences during the experiment. Korean speech-to-text services transcribed the recorded interviews.\footnote{\url{https://clovanote.naver.com}} Two lead authors independently open-coded each group's first three out of ten interviews, focusing on moderation experiences with \system{} and emotion change during the experiment. All authors then had discussion sessions to develop an initial codebook by discussing emerging themes, addressing inconsistencies, and resolving disagreements to reach a consensus. Based on the initial codebook, we coded the remaining interviews and had iterative discussion sessions to finalize the codebook.

\subsection{Ethical Considerations}
\label{sec:method:ethic}
Our institution's Institutional Review Board~(IRB) approved all study phases for ethical compliance. Each participant was pseudo-anonymized using a unique nickname throughout the research, including the user study, data analysis, and reporting. The participants were informed of their right to decline any interview questions. We notified participants in advance of the study that they would engage with hate speech content, which could impact their mental well-being. We also notified them that they could withdraw their participation at any time. % in the introductory session. 
\section{Results}
\label{sec:results}
We assess \system{}'s feasibility as a moderator support tool by answering the following RQs: (1)~How does each feature of \system{} contribute to moderators' mental well-being during hate speech moderation? (\S\ref{sec:results:wellbeing-quan}, \S\ref{sec:results:wellbeing-qual}) and (2)~How does each feature of \system{} influence moderation strategies and contribute to moderators' performance in hate speech moderation? (\S\ref{sec:results:performance-quan}, \S\ref{sec:results:performance-qual}) %To contextualize our findings, we first explain participants' main strategies for moderating hate speech using \system{} as observed during the study. %To understand \system{}'s role in supporting effective hate speech moderation while reducing emotional burden, we identified three main strategies participants used to moderate hate speech with \system{}'s style change features. 

We present an overview of quantitative findings for each research question and contextualize our results with insights from the semi-structured interviews.
% In advance of \system{}'s role in supporting moderators for effective hate speech moderation with less mental burden, we identified three main strategies to moderate hate speech with \system{}. 

% ====================================================================================================================================
\subsection{Quantitative Findings: \system{}'s Impact on Moderators' Mental Well-Being}
\label{sec:results:wellbeing-quan}
We quantitatively evaluate how each feature of \system{} impacts moderators' mental well-being during hate speech moderation~(RQ1). First, we analyze participant emotions and fatigue changes after the hate speech moderation task across different experimental conditions. We compare the perceived hate severity of comments modified by \system{} to that of unmodified comments. We then explore participants' self-reported perceptions of the effectiveness of \system{} in protecting mental well-being.
% We then examined how \system{} prevents the normalization of hateful and bias opinion, where hateful opinions might be perceived as acceptable, 
% Lastly, we assessed participants’ perceptions of \system{} as a protective tool for their mental well-being through perceived effectiveness score and qualitative insights.

\subsubsection{Impact of \system{} on Emotional State}

\begin{figure}[t]
    \centering
    \begin{minipage}{0.49\textwidth}
        \centering
        \includegraphics[width=\textwidth]{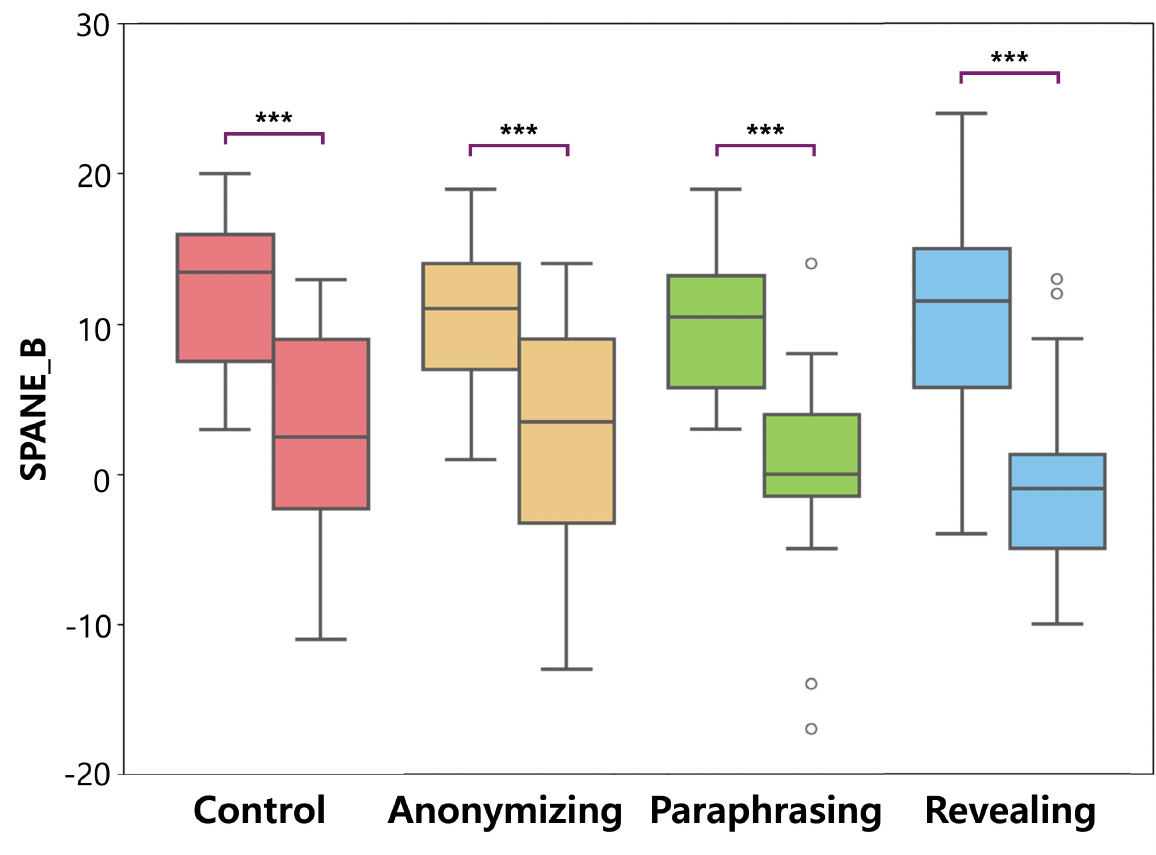}
        \vspace{-20pt}
        \caption{Changes in participants' SPANE\_B before and after the moderation task. *** represents $p$<.001 of Wilcoxon signed-rank test result.}
        \label{fig:spane_b}
    \end{minipage}%
    \hfill
    \begin{minipage}{0.49\textwidth}
        \centering
        \includegraphics[width=\textwidth]{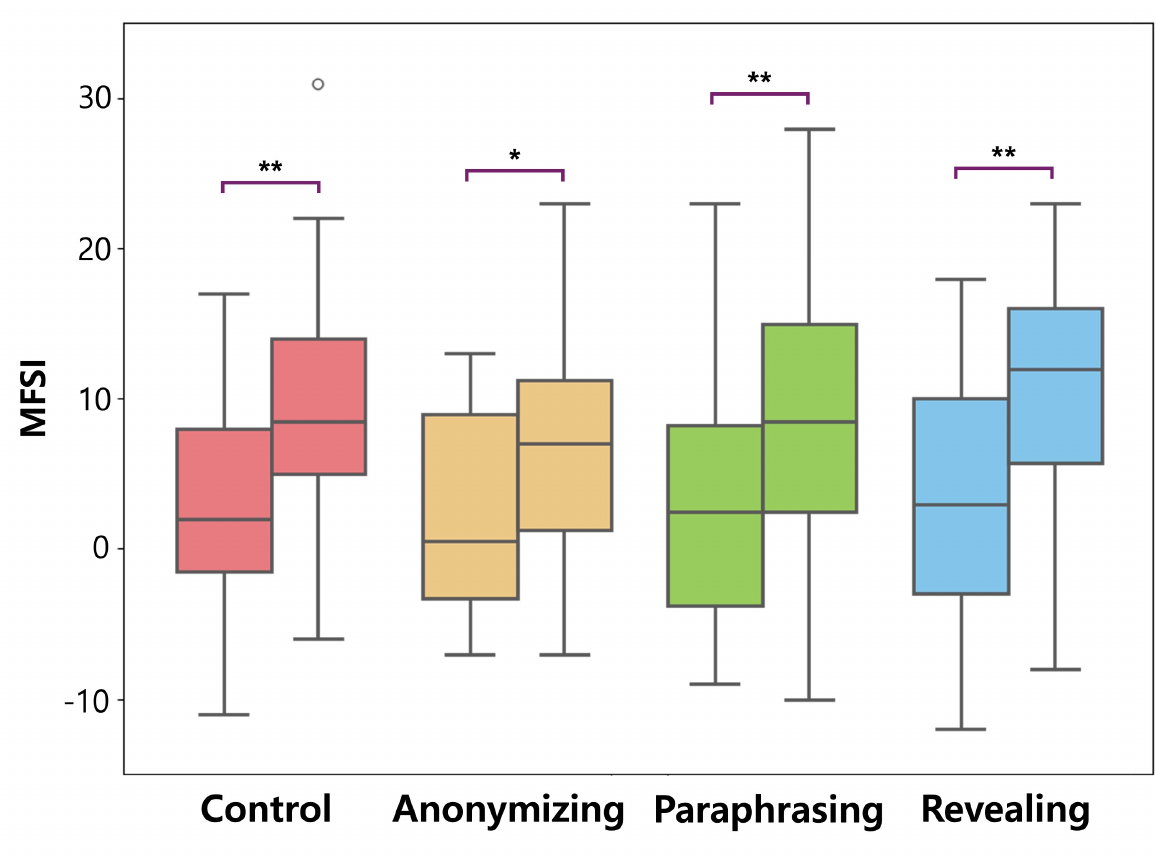}
        \vspace{-20pt}
        \caption{Changes in participants' \fatigue{} before and after the moderation task. * represents $p$<.05 and ** represents $p$<.01 of Wilcoxon signed-rank test result.}
        \label{fig:fatigue}
    \end{minipage}
    \vspace{-3pt}
\end{figure}

We first examined how \system{} influenced participants' emotional state, focusing on changes in SPANE\_B scores before and after the moderation task.
Fig.~\ref{fig:spane_b} shows each group's changes in SPANE\_B scores. The average SPANE\_B score in the pre-survey was 12.00 for the control group, 10.65 for the anonymizing group, 10.35 for the paraphrasing group, and 10.35 for the revealing group. A one-way ANOVA test found no statistically significant difference in SPANE\_B results across groups in the pre-survey~($F$=0.36, $p$=.785), indicating that participants started the moderation task with similar emotional states. 

In the post-survey, the average SPANE\_B score was 1.95 in the control group, 2.70 in the anonymizing group, 0.20 in the paraphrasing group, and -0.30 in the revealing group. A one-way ANOVA test indicates no statistically significant difference in SPANE\_B results across groups in the post-survey~($F$=0.84, $p$=.477). This result shows that although participants' emotional states were significantly more negative after the moderation task, this change was roughly equal across all experimental groups.
%However, we found that participants' emotions became significantly negative after the hate speech moderation based on within-subjects comparison of SPANE\_B across all groups. 
% group showed a significant decrease in SPANE\_B after the moderation tasks within-subjects, indicating that their emotion became significantly negative after the hate speech moderation across all groups. 
The descriptive statistics and Wilcoxon signed-rank test results are in Supplementary material~\ref{sec:supp:spaneb} and~\ref{sec:supp:spaneb_wilcoxon}, respectively.

\subsubsection{Impact of \system{} on Fatigue}
\label{sec:results:wellbeing:fatigue}
Next, we examined how \system{} affected participants' fatigue levels based on \fatigue{} scores. Fig.~\ref{fig:fatigue} shows the changes in participants' \fatigue{} scores before and after the moderation task for each group. The average \fatigue{} score in the pre-survey was 2.60 for the control group, 2.25 for the anonymizing group, 2.80 for the paraphrasing group, and 3.75 for the revealing group. A one-way ANOVA test revealed no statistically significant difference in \fatigue{} across groups in the pre-survey~($F$=0.36, $p$=.785), showing that participants started the moderation task with similar fatigue levels. 

In the post-survey, the average \fatigue{} score was 9.05 in the control group, 6.20 in the anonymizing group, 8.35 in the paraphrasing group, and 10.50 in the revealing group. The post-survey one-way ANOVA test indicated no statistically significant difference in fatigue scales across groups~($F$=0.90, $p$=.447). As with the SPANE\_B results we analyzed previously, we found that participants in all groups felt significantly greater fatigue after the hate speech moderation based on within-subjects comparison of \fatigue{}, but no experimental condition mitigated this increase in fatigue. 
% \fatigue{} significantly increased after the moderation task. 
We report the descriptive statistics and Wilcoxon signed-rank test results in Supplementary material~\ref{sec:supp:fatigue} and~\ref{sec:supp:fatigue_wilcoxon}, respectively.

In summary, both emotional and fatigue scales showed similar trends: participants felt more negative emotions and greater fatigue after performing hate speech moderation, regardless of the assigned study group. 
% with emotional scales indicating a decrease in positive affect and fatigue scales showing an increase in fatigue, without significant differences between groups.

\subsubsection{Perceived Hate Severity}
\label{sec:results:wellbeing:severity}

\begin{figure}[t]
    \centering
    \includegraphics[width=0.5\textwidth]{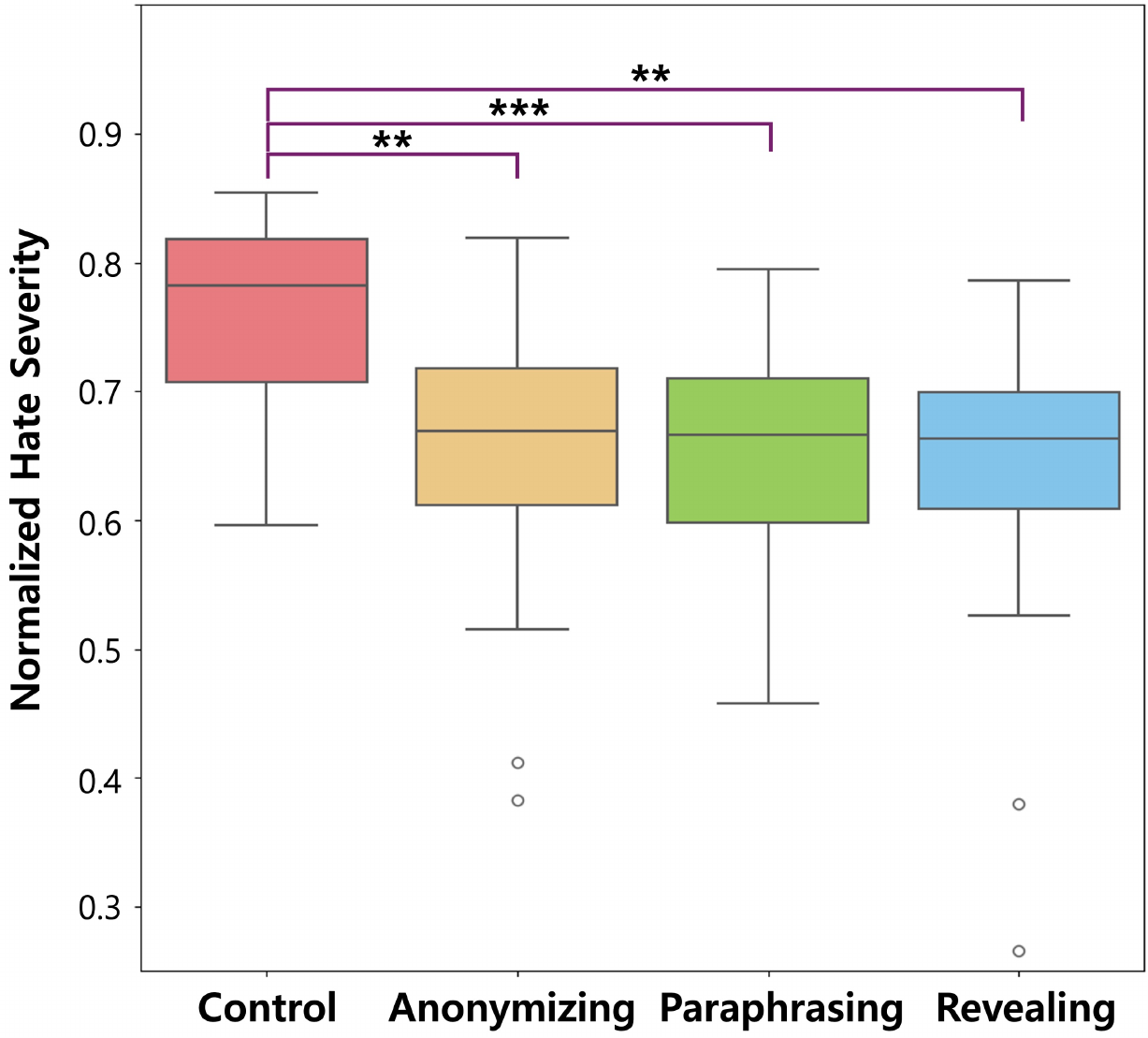}
    \vspace{-5pt}
    \caption{Comparison of normalized hate severity scores of 50 hate speech comments across four groups. ** indicates $p$<.01 and *** indicates $p$<.001. %Normalized Hate Severity in comments across different groups. \textit{“How severe is the hate speech perceived in the comment?”} (1: Not at all hateful, 5: Very hateful)
    }
    \label{fig:severity}
    % \vspace{-5pt}
\end{figure}

Following our assessment of emotion and fatigue, we examined how participants rated the severity of hate in the comments across all groups. We found that the control group rated the severity of hate in the 50 hate speech comments as significantly higher~(mean=3.94,~std=0.58) than the anonymizing group~(mean=3.56,~std=0.75), the paraphrasing group~(mean=3.66,~std=0.65), and the revealing group~(mean=3.58,~std=0.80). To clearly observe the distribution of perceived hate severity across groups, independent of individual bias, we applied z-score normalization to the perceived hate severity scores of all 100 comments for each participant~(Fig.~\ref{fig:severity}). A one-way ANOVA test found a statistically significant difference in the normalized hate severity distribution across groups~($F$=6.53, $p$=.001). Pair-wise two-tailed t-tests revealed that the three experimental groups evaluated the comments as significantly less severe than the control group: the anonymizing group~($U$=3.54,~$p$=.006), the paraphrasing group~($U$=4.27,~$p$=.001), and the revealing group~($U$=3.94,~$p$=.002). There was no significant difference among the experimental groups (i.e., anonymizing, paraphrasing, and revealing groups). This result indicates that \offensiveparaphrasing{} and \targetanony{} reduce the perceived hatefulness of hate speech.
% This result indicates that \offensiveparaphrasing{} reduces the perceived hatefulness of hate speech the most, with \targetanony{} also contributing to some extent.
These effects appear to persist in \system{}, even when participants had the option to view the original expressions through \targetrevealing{} and \offensiverevealing{}. As participants perceived comments as less severe with \system{}, we next examine how these perceptions contributed to \system{}'s perceived effectiveness in protecting mental well-being.

\subsubsection{Perceived Effectiveness of \system{} in Protecting Mental Well-Being}

\begin{figure}[t]
    \centering
    \includegraphics[width=\textwidth]{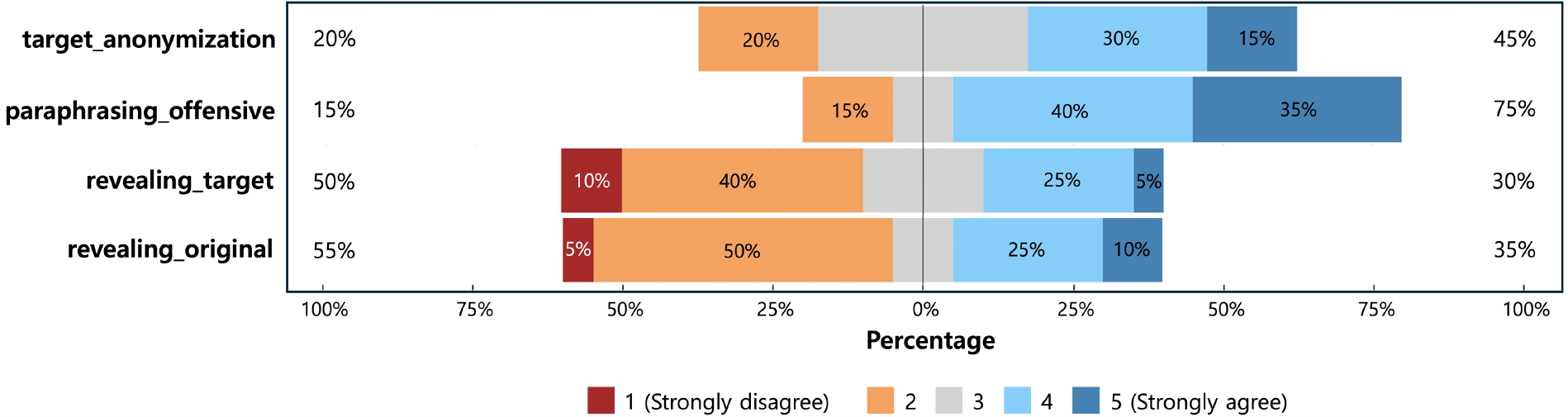}
    \vspace{-20pt}
    \caption{Distribution of impact on the mental well-being of each feature%~(i.e.,~\targetanony{}, \offensiveparaphrasing{}, \targetrevealing{}, and \offensiverevealing{})
    . Responses range from 1 (Strongly Disagree) to 5 (Strongly Agree).}
    \label{fig:emotional_wellbeing}
    \vspace{-5pt}
\end{figure}

For the final quantitative analysis for RQ1, Fig.~\ref{fig:emotional_wellbeing} illustrates the distribution of the perceived effectiveness of \system{}'s features for protecting mental well-being based on analysis of a five-point Likert scale survey question. 
Participants perceived \targetanony{}'s effectiveness in protecting mental well-being as moderate: 45\% of participants agreed or strongly agreed that \targetanony{} was effective in protecting their mental well-being, while 20\% disagreed. 
Participants' perceptions of the effectiveness of \offensiveparaphrasing{} in protecting them from hate speech were more positive, with 75\% agreeing that it offered emotional protection. 
% In the target group, 45\% of participants at least agreed that \targetanony{} was effective in protecting their mental well-being, while 20\% of them disagreed. The offensive group was more positive toward \offensiveparaphrasing{}'s role in mental well-being protection, showing 75\% at least agree that \offensiveparaphrasing{} protected them. On the other hand, the effectiveness of \targetrevealing{} and \offensiverevealing{} on mental well-being protection was controversial. While 30\% of participants from the revealing group evaluated \targetrevealing{} as effective in protecting their mental well-being, 50\% did not. Similarly, 35\% of them evaluated \offensiverevealing{} as effective in protecting their mental well-being, but 55\% disagreed. We report the factors contributing to these perceptions of \system{} in the later section.

In contrast to the positive evaluations of \targetanony{} and \offensiveparaphrasing{}, participants had mixed evaluations of the effectiveness of \targetrevealing{} and \offensiverevealing{} features. While 30\% of participants from the revealing group evaluated \targetrevealing{} as effective in protecting their mental well-being, 50\% did not. Similarly, 35\% evaluated \offensiverevealing{} as effective in protecting their mental well-being, but 55\% disagreed. 
These results indicate that although \targetanony{} and \offensiveparaphrasing{} are largely perceived as supportive of mental well-being, \targetrevealing{} and \offensiverevealing{} had mixed evaluations.

% To sum up, all experimental groups rated the altered comments as less severe hate speech compared to the control group, aligning with the high perceived effectiveness of the \targetanony{} and \offensiveparaphrasing{} in protecting mental well-being. This protection also led to fewer changes in negative emotions after the moderation task with \system{}. In the interview, participants also appreciated \targetanony{} and \offensiveparaphrasing{} for preventing their personal values from being influenced by hate speech. Overall, we observed that \targetanony{} and \offensiveparaphrasing{} protected participants' emotions by concealing original expressions (i.e.,~lowering perceived severity levels). However, we posit that positive friction features, \targetrevealing{} and \offensiverevealing{}, possibly have increased the moderation workload (e.g., requiring more time for moderation), which may offset the potential benefits on mental well-being. We will discuss the impact of mental well-being on \system{} in the discussion section later. 
Overall, participants felt more negative emotions and greater fatigue after moderating hate speech, regardless of the study condition. However, the three experimental groups rated the severity of hate speech in the comments they saw as less severe than the control group, suggesting that anonymizing targets and paraphrasing offensive expressions can make hate speech appear less hateful.

We now turn to qualitative insights to complement these mixed quantitative findings, exploring participants' experiences and perceptions of \system{}'s features in supporting their mental well-being. We also discuss potential explanations for these mixed results in \S~\ref{sec:discussion:discrepancy}.

% ====================================================================================================================================
\subsection{Qualitative Insights: \system{}'s Impact on Moderators' Mental Well-Being}
\label{sec:results:wellbeing-qual}
Through qualitative analysis, we gain deeper insights into participants' experiences with \system{}'s features and their perceptions of its role in supporting mental well-being, including its effectiveness in preventing the normalization of hateful and biased opinions.

\subsubsection{Aspects of \system{} that Were Perceived as Helpful for Mental Well-Being}
% What makes participants felt supportive from modified hate speech 
% Factors That Made \system{} Perceived Metally Supporting
% Mental wellbeing support factors in \system{}
% Perceived Mental Well-Being Supportive Aspects of \system{}
In the interview, participants shared diverse perspectives on the value and impact of each feature.
Participants expressed varying views on \targetanony{}'s value. While some were \textbf{uncertain about anonymization's effectiveness}, others highlighted how it \textbf{helped when the participant shared an identity} with the target. For example, A01 noted, \textit{``When there are hate speech comments directed at a certain gender, if I belong to that gender, it could hurt me more and make me feel more depressed than when the target is hidden.''}
% While some participants were not confident in the value of the feature, others highlighted how it helped in cases where the participant shared an identity with the target. For example, according to A01, \textit{``when there are hate speech comments directed at a certain gender, if I belong to that gender it could hurt me more and make me feel more depressed compared to when the target is hidden.''} -- A01.
% A07 noted, ``I think the impact on my mental well-being would be similar. Basically, there are so many negative words or discriminatory remarks that I still end up feeling negative emotions.'' % 심리 상태에 미치는 영향은 비슷할 것 같습니다. 그냥 기본적으로 뭔가 부정적인 단어나 좀 차별적인 발언이 굉장히 많아서 여전히 부정적인 감정이 생기는 것은 동일할 것 같습니다.(신바람나는 표범, target) 
 % 예를 들어서 특정 성별에 대해서 그런 비난 발언을 했을 때 제가 그 성별이면 또 그걸 더 가려져 있을 때보다 더 상처받을 수도 있고 더 우울해질 수도 있고 그러니까 (정다운북극곰, target)
In addition, some participants described \targetanony{} as effectively preventing emotional contagion from pejorative terms. As A03 explained, \textit{``I think hiding [targets] is much better for my emotions. If I'm repeatedly exposed to unpleasant words like `Feminazi' [pejorative term for feminist] or `Democrap' [pejorative term for Democratic Party], I feel like my emotions might get influenced by them.''}  % 제 생각에 회색으로 가리고 하는 게 훨씬 제 감정에 더 나은 것 같은 게 뭐 '페미'도 있고 '더불어적폐당 국회의원' 이렇게 약간 좋지 않은 단어들에 계속 노출되면 저 또한 감정이 거기에 동화될 것 같기도 하고... (미소짓는북극곰, target)

For \offensiveparaphrasing{}, participants reported \textbf{a noticeable impact on emotional well-being}. P01 described, \textit{``When I read the softened expressions, I don't really feel offended.''} % 완화된 표현을 읽었을 때는 크게 불쾌하다는 생각은 일단 막 들지는 않는 것 같아요. 직접적으로 비속어를 읽으면 훨씬 더 심리 상태에 안 좋은 영향을 주는 것 같습니다. (느긋한고슴도치, offensive)
Some participants emphasized that, although they could infer the original expressions, the protection provided by the \offensiveparaphrasing{} was still meaningful. P02 said \textit{``Imagining is just imagining, so I told myself that I was over-interpreting the comments. Because of that, I don't think [moderating the hate speech] really had much of an emotional impact on me.''}
% 우선 상상하는 거는 그냥 상상일 뿐이기 때문에 이제 내가 과대 해석하는 거다라고 생각하고 댓글을 봤기 때문에 별로 감정적으로 타격은 없었던 것 같고 (점잖은기린, offensive)

When it came to the revealing features, some participants described how \textbf{encountering the target of hate speech was surprising}, saying \textit{``When the target was hidden, it felt like vague or meaningless talk, but once I clicked and saw the real subject, it felt more emphasized and dramatic. It didn’t make me feel worse or sad, but I was just surprised''} (R02). % 그냥 타깃이 없을 때는 뜬구름 잡는 얘기를 하는 것처럼 느껴졌는데 막상 클릭을 해서 진정한 주체를 딱 보면 좀 강조가 돼서 드라마틱하게 느껴진다고 해야 되나 ... 그렇다고 막 감정이 나빠지거나 아니면 슬퍼지거나 이런 건 아닌데 그냥 놀랐던 것 같아요. (정다운개구리, revealing)
Participants also mentioned that inferring the original expressions and reading the original felt a bit different. R07 noted \textit{``I had expected similar expressions to some extent, but there were still moments when I felt some discomfort. But it wasn’t like my mental state was severely affected.''} % 약간 비슷하게 어느 정도 (공격적인 표현을) 예상은 했지만 좀 불쾌감이 느껴지는 경우가 좀 있었던 것 같아요. 그런데 원본을 봤을 때 그냥 기분이 나빠졌다지, 너무 막 멘탈이 힘들고 이런 건 없었던 것 같아요. (신바람나는팬더, revealing)
These reactions suggest that exposure to the original target and offensive expressions was the source of discomfort, and the hiding features gave participants control over whether to experience this.
% These reactions suggest that exposure to the original target and offensive expressions is naturally displeasing.
% These reactions suggest that it was the exposure to the original target and offensive expressions, rather than the revealing action itself, that made participants perceive \targetrevealing{} and \offensiverevealing{} as less effective in protecting their mental well-being. % 원본 콘텐츠에 노출되는 것 자체는 당연히 사람들이 부정적인 영향을 받았는데 피험자가 원할 때 선택적으로 확인햇기 때문에 마음의 준비나 예상 같은 걸 통해 처음부터 원본에 노출된 것(control)보다는 비교적 영향이 덜했다.

Some participants explicitly described how they valued this type of control. They explained how \targetrevealing{} and \offensiverevealing{} enabled a \textbf{phased approach that made them feel less offended} by hate speech. R08 explained how \textit{``Going through the somewhat complicated process of checking the original made me feel less offended. So, whether the target was directed at me or my group, I think this more complex process helped reduce the impact on my mental state.''} % 좀 복잡하지만 그 원본을 확인하는 과정을 거치다 보니까 좀 덜 공격적으로 다가오는 것 같아요. 그래서 그게 이제 저를 포함한 집단에 대한 타겟이든 아니든 간에 조금 복잡한 과정을 거치지만 좀 멘탈 영향이 덜 가지 않나 생각을 하고요. (신바람나는나비, revealing)
R04 also described how revealing allowed them time to prepare mentally, saying \textit{``For any content that involves hate speech, I feel like I need time to mentally prepare before clicking to reveal it. If I were a moderator, I would probably encounter mostly hate speech, so instead of trying to encounter all the content right away, I think I’d prefer to know first that it’s likely to contain violent or offensive language so I can be mentally prepared.''} 
% 그러니까 혐오 표현에 해당하는 모든 내용은 약간 내가 이걸 클릭을 마음의 준비를 할 시간이 필요한 것 같아요. 제가 만약에 댓글 관리자라면 대부분 접하는 게 사실 혐오 표현의 댓글일 텐데 댓글의 내용을 모두 이해하는 것보다는 저는 이거는 폭력성이 좀 노출될 수 있는 댓글임을 먼저 알고 마음의 준비를 하는 것을 먼저 원할 것 같아요. (귀여운토끼, revealing)

To sum up, some participants noted how \targetanony{} and \offensiveparaphrasing{} could be effective in preventing their attitudes from being influenced by hate speech, and others noted how \targetrevealing{} and \offensiverevealing{} worked as a form of positive friction. We next examine how these perceptions influenced their attitudes during the moderation process.

\subsubsection{\system{} Safeguarded Against Normalization of Biased and Hateful Opinions}
% In this study, the experimental group did not experience this normalization effect as strongly, thanks to specific features built into the system. These system features likely provided some form of intervention that counteracted the typical desensitization or acceptance process. Such features might include:
% Through these design interventions, participants in the experimental group felt that the system helped prevent the normalization effect, making hateful opinions less likely to be perceived as acceptable or normative.
Participants in the control group reported concerns that exposure to hate speech not only triggered emotional distress but also could lead to \textbf{normalization of biased and hateful opinions from the comments}.
%contributing to a phenomenon they described as `bias contagion.' 
% Since I kept seeing mostly negative things, I started thinking, `Is the world really this full of negativity?' Also, 
Per C07, \textit{``... even though I don't personally think the baseball players did anything wrong, reading the comments made me wonder, `Did they really do something wrong?' ... It seems like these comments make me think more negatively, and I don't feel good about that.''} % 계속 대부분 부정적인 것만 보다 보니까 이제 그냥 좀 세상에 이렇게 약간 부정적인 일들이 많나 약간 생각도 들고, 그리고 딱히 저는 야구 선수들이 잘못된 행동을 했다고 생각하지 않는데 (댓글 보니) `잘못했나?' 약간 이렇게 생각이 들기도 하고 ... 이런걸 부정적으로 생각하게 되는 계기가 되는 것 같아서 조금 안 좋은 것 같습니다. (미소짓는부엉이, control)
% repetitive exposure to negative contents led participants to feel that their attitudes toward specific groups or topics were being subtly shaped negatively.

In contrast, the experimental groups felt that anonymizing specific targets of hate speech or paraphrasing offensive expressions \textbf{safeguarded them from the normalization of biased or hateful opinions} from hate speech, creating a needed distance between their perspectives and the negative opinions they encountered during the experiment. 
For instance, A01 reflected on the role of \targetanony{} in containing negative emotional contagion: \textit{``When it's hidden, it feels like I'm just dealing with my own negative emotions [about the comments]. If I read comments about a specific country without any hiding, I end up reading all the criticisms directed at that country. This makes me feel like I might start to resonate with those negative emotions and end up disliking that country as well.''}
% 가리는 게 좋을 것 그러니까 분명 이제 가려도 많이 부정적인 감정이 들긴 하는데 안 가렸을 때는 그 부정적인 감정이 특정 집단에 쏠릴 수도 있을 것 같아요. 그러니까 가렸을 때는 그냥 저 혼자 부정적인 감정이 드는 느낌인데 약간 예를 들어서 제가 만약에 되게 특정 나라에 대한 그런 가리지 않은 상태에서 보면은 그 특정 나라에 대한 비난들을 다 제가 읽게 되는 거잖아요. 그래서 뭔가 그 특정 나라에 저도 그 감정이 동화되어서 저도 그 나라를 싫어하게 될 것 같은 느낌이라고 해야 되나 (정다운북극곰, target)
Similarly, P05 reflected on the experience of reading hate speech in the recruitment form and said, \textit{``When I read hate speech in the recruitment form, I had the feeling that if I kept seeing these kinds of comments, I might start to think like the people who wrote them. But today, I didn’t feel like my emotions became extremely negative, and I was less worried that doing this kind of work continuously would negatively affect my mental state.''} % % 처음에 이제 지원할 때 받았던 그 구글 폼을 작성할 때는 이거 계속 보면은 내가 정말 이런 사고를 하는 사람이랑 비슷해질 것 같다 이런 느낌을 받았었는데. 오늘 같은 경우는 감정이 엄청 부정적으로 변하거나 '앞으로 내가 이런 거를 작업을 계속했을 때 내 정신 세계가 이상하게 변할 것 같다' 그런 걱정은 덜 하게 됐던 것 같아요. (팔팔한곰, offensive)
% These reflections suggest that hate speech moderation led to concerns about normalization of biased and hateful opinion, while \system{} helped protect participants from the negative effects of hateful opinions on their perspectives.
These reflections suggest that, while reviewing hate speech can influence participants' attitudes, \system{} offered a protective layer for moderators’ mental well-being by supporting them in maintaining emotional distance from biased and hateful opinions. 
% These reflections suggest that \system{} helped protect participants from the negative impact of hate speech on their \jh{perspectives}.

% ============================================================================================================================================
\subsection{Quantitative Findings: \system{}'s Impact on Moderation Performance}
\label{sec:results:performance-quan}

We explore how altering comment text through the features of \system{} affects moderation performance~(RQ2) %. We assessed \system{}’s impact on moderation performance 
by analyzing moderation accuracy and recall, task completion time, and perceived effectiveness for hate speech moderation. 

\subsubsection{Moderation Accuracy and Recall}

\begin{table}[t]
\caption{Descriptive analysis of moderation accuracy and recall.}
\label{tab:acc}
\resizebox{\textwidth}{!}{%
\begin{tabular}{ccccccclcccccc}
\Xhline{2\arrayrulewidth}
\multirow{3}{*}{\textbf{Group}} & \multicolumn{6}{c}{\textbf{Moderation accuracy}}                                                                                       &  & \multicolumn{6}{c}{\textbf{Moderation recall}}                                                                                    \\ \cline{2-7} \cline{9-14} 
\multicolumn{1}{l}{} & \multirow{2}{*}{\textbf{Mean}} & \multirow{2}{*}{\textbf{Std}} & \multirow{2}{*}{\textbf{Min}} & \multirow{2}{*}{\textbf{Max}} & \multicolumn{2}{c}{\textbf{Shapiro-Wilk}} &  & \multirow{2}{*}{\textbf{Mean}} & \multirow{2}{*}{\textbf{Std}} & \multirow{2}{*}{\textbf{Min}} & \multirow{2}{*}{\textbf{Max}} & \multicolumn{2}{c}{\textbf{Shapiro-Wilk}} \\ \cline{6-7} \cline{13-14} 
\multicolumn{1}{l}{} &                       &                      &                      &                      & \textbf{W}             & \textbf{p-value}          &  &                       &                      &                      &                      & \textbf{W}             & \textbf{p-value}          \\ \cline{1-7} \cline{9-14} 
Control              & 0.79                  & 0.12                 & 0.54                 & 0.96                 & 0.97          & .749             &  & 0.62                  & 0.28                 & 0.05                 & 0.98                 & 0.95          & .375             \\
Anonymizing               & 0.75                  & 0.13                 & 0.52                 & 0.92                 & 0.94          & .212             &  & 0.54                  & 0.31                 & 0.01                 & 0.94                 & 0.92          & .089             \\
Paraphrasing            & 0.80                  & 0.09                 & 0.68                 & 0.91                 & 0.97          & .851             &  & 0.67                  & 0.24                 & 0.34                 & 0.97                 & 0.95          & .299             \\
Revealing            & 0.79                  & 0.11                 & 0.59                 & 0.92                 & 0.97          & .752             &  & 0.71                  & 0.26                 & 0.27                 & 1.00                 & 0.94          & .250            \\\Xhline{2\arrayrulewidth}
\end{tabular}%
}
\end{table}

Table~\ref{tab:acc} presents a descriptive analysis of each group's moderation accuracy and recall. Overall, the average accuracy scores were similar across all groups. The average accuracy was 0.79 for the control group, 0.75 for the anonymizing group, 0.80 for the paraphrasing group, and 0.79 for the revealing group. A one-way ANOVA test was conducted, as the Shapiro-Wilk test confirmed a normal accuracy distribution in each group, and no significant differences were found across the groups ($H$=1.05,~$p$=.377). 
We also investigated moderation recall, the ratio of deleted hate speech comments to the ground truth hate speech. The average recall was 0.62 for the control group, 0.54 for the anonymizing group, 0.67 for the paraphrasing group, and 0.71 for the revealing group. A one-way ANOVA test was conducted, as the Shapiro-Wilk test confirmed a normal distribution of recall in each group, and no significant differences were found across the groups ($H$=1.71,~$p$=.173).

%On the other hand, the paraphrasing group demonstrated higher \jh{recall} than the control group, despite the offensive expressions being paraphrased as less aggressive. 
To summarize, there was no statistically significant difference in moderation accuracy and recall across groups despite \targetanony{} and \offensiveparaphrasing{} limiting the information participants received about the comments. Still, the paraphrasing group demonstrated higher moderation recall than the control group, despite the offensive expressions being paraphrased as less aggressive. The revealing group achieved the highest moderation recall. We explore what enables participants to moderate accurately and sensitively despite content modification in \S~\ref{sec:results:performance:acc-sens-qual}.
%by sensing the offensive nuance and inferring the original expressions from the paraphrased ones. Furthermore, the revealing group achieved the highest moderation \jh{recall} among groups, benefiting from \offensiveparaphrasing{} at first and effectively moderating comments with pejorative terms by revealing features. 
% While the target group showed lower moderation accuracy and \jh{recall} compared to the control group,

% [offensive_paraphrasing] Could make objective moderation decisions by paraphrasing pejorative terms
% [target_hiding] Supported objective moderation decision-making

\subsubsection{Task Completion Time}
%     \centering
%     \includegraphics[width=\textwidth]{fig/plots/plot_time.png} % Replace with your figure file
%     \caption{Task completion time by moderation step.}
%     \label{fig:time}
% \usepackage{multirow}

\begin{figure}[t]
    \centering
    \begin{minipage}[]{0.48\linewidth}
        \centering
        \includegraphics[width=\linewidth]{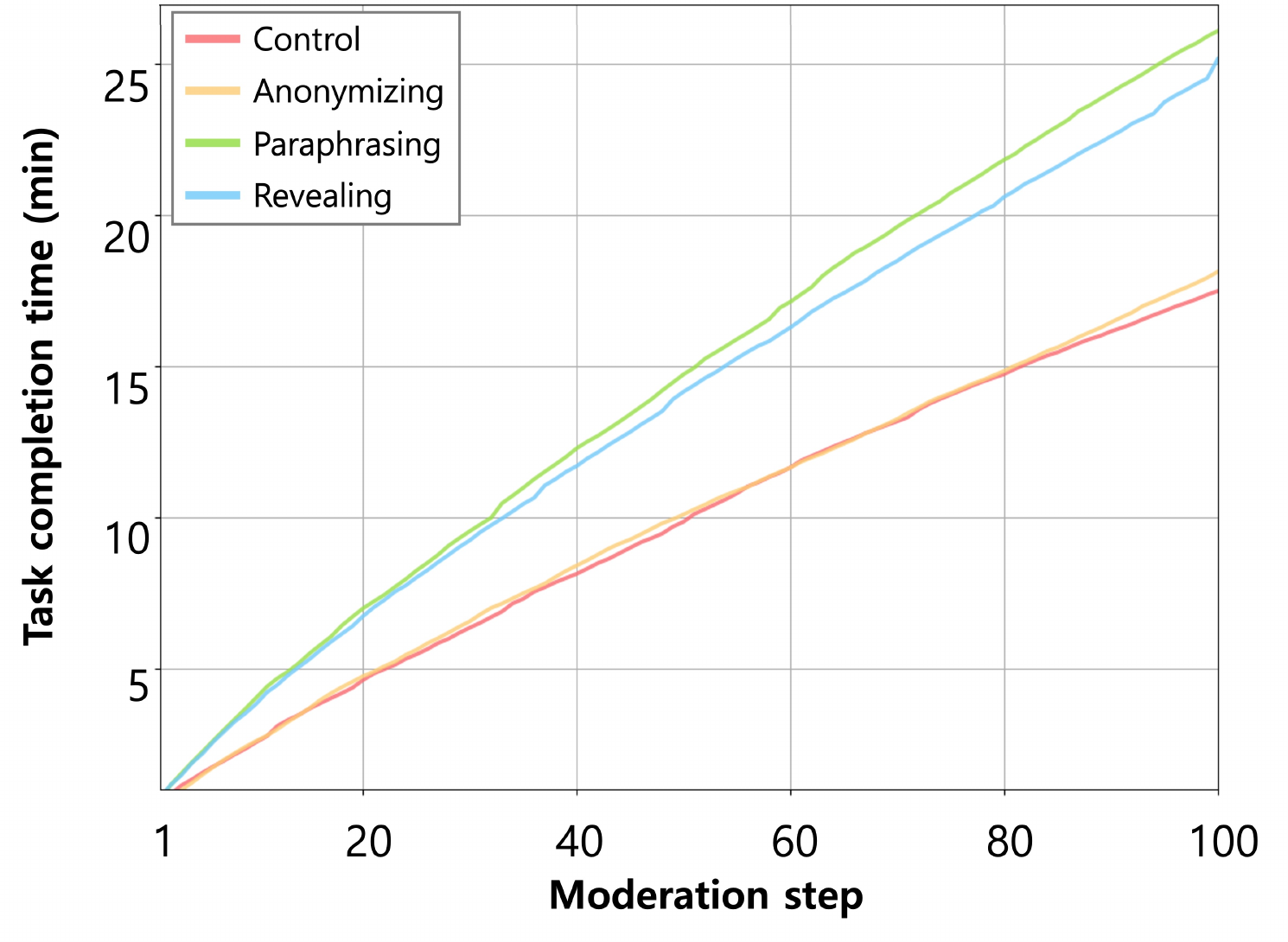}
        \vspace{-20pt}
        \caption{Cumulative task completion time by moderation step.}
        \label{fig:time}
    \end{minipage}%
    \hfill
    \begin{minipage}[]{0.51\linewidth}
        \centering
        \captionof{table}{Descriptive analysis of task completion time in minutes. * indicates $p$<.05.}
        \label{tab:time}
        \resizebox{\textwidth}{!}{%
        \begin{tabular}{ccccccc}
        \Xhline{2\arrayrulewidth}
        \multirow{2}{*}{\textbf{Group}} & \multirow{2}{*}{\textbf{Mean}} & \multirow{2}{*}{\textbf{Std}} & \multirow{2}{*}{\textbf{Min}} & \multirow{2}{*}{\textbf{Max}} & \multicolumn{2}{c}{\textbf{Shapiro-Wilk}} \\ \cline{6-7} 
         &                       &                      &                      &                      & \textbf{W}             & \textbf{p-value}          \\ \hline
        Control              & 18.07                 & 5.08                 & 10.45                & 29.13                & 0.96          & .125             \\
        Anonymizing               & 18.13                 & 7.30                 & 9.60                 & 34.95                & 0.87          & .010$^{*}$         \\
        Paraphrasing            & 26.10                 & 9.15                 & 12.12                & 48.47                & 0.95          & .326             \\
        Revealing            & 25.17                 & 9.66                 & 10.72                & 48.12                & 0.96          & .630            \\ \Xhline{2\arrayrulewidth}
        \end{tabular}%
        }
    \end{minipage}
\end{figure}

Fig.~\ref{fig:time} plots the cumulative task completion time across moderation tasks, and Table~\ref{tab:time} provides descriptive statistics for the task completion time of each group. Overall, the control and anonymizing groups displayed a similar trend, completing the moderation of 100 comments in an average of 18.07 minutes and 18.13 minutes, respectively. The paraphrasing group took the longest to complete, averaging 26.10 minutes. While the revealing group showed a similar trend to the paraphrasing group, they completed the task slightly faster, with an average time of 25.17 minutes, despite the availability of more interactive features. 
% KruskalResult(statistic=15.38364764762421, pvalue=0.0015164839475359363)

% Mann-Whitney U test
Kruskal-Wallis test revealed a statistically significant difference across the groups ($H$=15.38,~$p$=.002), indicating that at least one group differs significantly from the others. There were significant differences in task completion time between the paraphrasing and control groups ($U$=84.5,~$p$=.002), paraphrasing and anonymizing groups ($U$=94.5,~$p$=.005), revealing and control groups ($U$=106.0,~$p$=.011), and revealing and anonymizing groups ($U$=110.0,~$p$=.015). We discuss the friction introduced by \system{}, which led to increased task completion times in hate speech moderation, in \S\ref{sec:discussion:text}.
% \subin{However, the gap between the two slower groups and the others was less than ten minutes for moderating 100 comments. }
% , which suggests that the additional efforts required to use \system{} (e.g.,~inferring the original, interacting with \targetrevealing{} or \offensiverevealing{}) did not dramatically slow down task performance. 
% approximately 30.7\% (control vs. revealing)

\subsubsection{Perceived Effectiveness for Hate Speech Moderation}
\label{sec:results:performance:effectiveness}

\begin{figure}[t]
    \centering
    \includegraphics[width=0.8\linewidth]{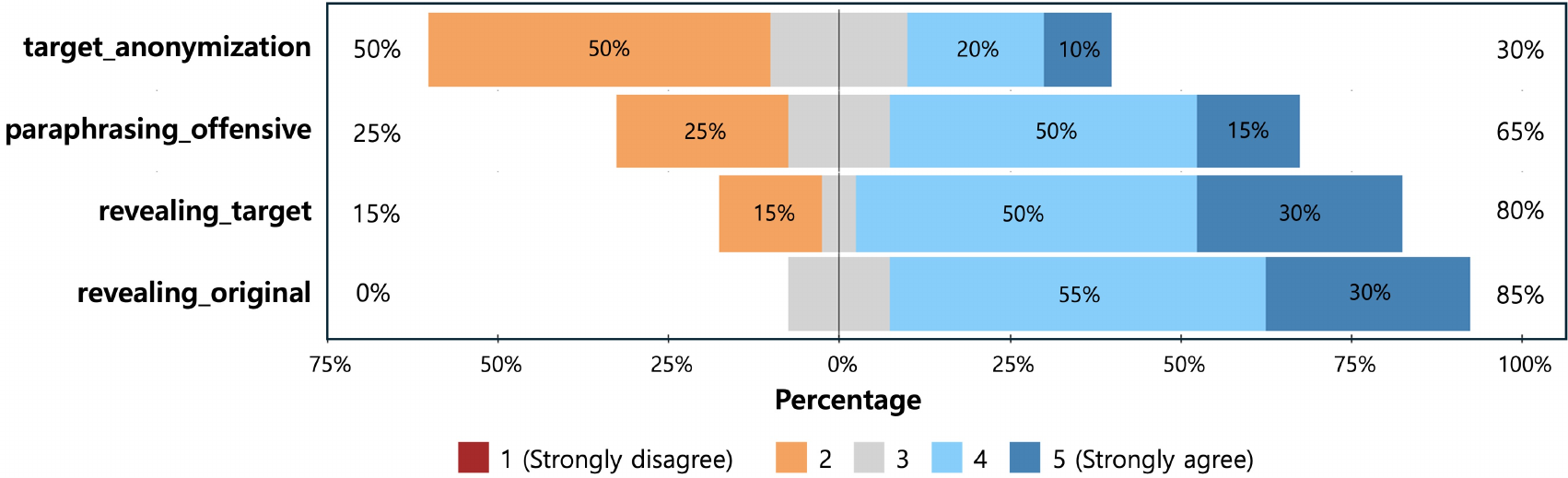}
    \vspace{-5pt}
    \caption{Distribution of perceived effectiveness of each feature %(i.e.,~\targetanony{}, \offensiveparaphrasing{}, \targetrevealing{}, and \offensiverevealing{}) 
    for hate speech moderation. Responses range from 1~(Strongly disagree) to 5~(Strongly agree).}
    \label{fig:perceived_performance}
\end{figure}
% target: 2(10), 3(4), 4(4), 5(2)
% offensive: 2(5), 3(3), 4(9), 5(3)
% target revealing: 2(3), 3(1), 4(10), 5(6)
% offensive revealing: 3(3), 4(11), 5(6)

% perceived clarity.
Fig.~\ref{fig:perceived_performance} illustrates the distribution of perceived effectiveness for each \system{} feature in hate speech moderation. In the anonymizing group, 30\% of participants agreed or strongly agreed that \targetanony{} was helpful for the moderation task, while 50\% disagreed. In the paraphrasing group, 65\% of participants agreed or strongly agreed that \offensiveparaphrasing{} helped moderate hate speech, while 25\% disagreed. In the revealing group, 80\% of participants agreed or strongly agreed that \targetrevealing{} was helpful for the moderation task, while 15\% disagreed. Finally, 85\% of the participants agreed or strongly agreed that \offensiverevealing{} was beneficial for performing moderation tasks, and none disagreed.

To summarize, we explored the impact of \system{} on moderation performance and found that, despite \targetanony{} and \offensiveparaphrasing{} modifying the original comments, overall accuracy was not negatively affected. Notably, both the paraphrasing and revealing groups demonstrated increased moderation recall, with \offensiveparaphrasing{} providing subtle cues in paraphrased expressions. 
For the task completion time, the paraphrasing and revealing groups took longer than the others.
Overall, participants considered \offensiveparaphrasing{}, \targetrevealing{}, and \offensiverevealing{} features helpful for moderating hate speech, despite \offensiveparaphrasing{} modifying the original expressions. Building on these quantitative findings, we next explore qualitative insights into participants' experiences and perceptions of how \system{}'s features support moderation performance.

% ============================================================================================================================================
\subsection{Qualitative Insights: \system{}'s Impact on Moderation Performance}
\label{sec:results:performance-qual}
In this section, we explore participants' strategies in applying \system{}'s features and examine how these approaches influenced moderation accuracy and recall, shedding light on the benefits and challenges of hate speech moderation with textual content modifications. 
% the challenges and complexities of hate speech moderation.

\subsubsection{Strategies to Moderate Hate Speech with \system{}}
\label{sec:results:performance:strategies}
% explicit한 거 보는거, overall content 보는거 => pejorative term 볼 수 있었다고 한 경우에는 perceived effectiveness?
% revealing feature selectively 쓰는거 => 1) moderation을 위해 쓴다 2) protection을 위해 쓴다 (=> well-being seciton의 perceived hate speech에서 revealing group의 hate severity가 control과 유의미한 차이가 나는 이유로 넣기)
% 아니면 impact on attitudes에서 푸는 게 좋을까? hmm hmm 

We first describe participants' strategies for moderating hate speech using \system{}: focusing on explicit, offensive language, evaluating overall content, and selectively revealing the original expressions. The revealing group varied in their use of revealing features, balancing accuracy in moderation with protecting their emotions.

%\noindent\textbf{Paying Attention to Explicit Offensive Expressions.} 
In the study groups where original offensive expressions were visible (i.e.,~control, anonymizing, and revealing groups), participants frequently \textbf{relied on the presence of explicit expressions} to determine whether a comment constituted hate speech. Many participants reported removing comments with raw and acute expressions. A01 noted, \textit{``I believe that even if the content is the same, the choice of words can determine whether a comment should be kept. ... Some people phrase things more gently, in a way that helps other people improve, while others use provocative words to hurt someone intentionally, and that's the real issue. ... So, I focused on the underlined words [offensive expressions] to see whether they used hateful language.''} % [Removed comments with raw and acute expression/criticism] 같은 내용을 말하더라도 어떤 단어를 쓰냐에 따라서 그 댓글의 유지 여부가 달라진다고 저는 생각하거든요. ... 누구는 좀 더 순화해서 최대한 그 사람이 고쳐지는 방향이 될 수 있도록 좋게 좋게 말하는 반면에, 어떤 사람들은 자극적인 단어로 그 사람이 상처 입히려하고, 그게 문제가 되는 거기 때문에 같은 내용이라도 단어 선택이 중요하다고 생각해서 그런 혐오적인 단어를 썼나 안 썼나를 보려고 밑줄 친 단어(offensive expressions)를 집중적으로 봤던 것 같아요. (정다운북극곰, target)

In particular, some participants mentioned removing comments containing pejoratives, as these words tended to ridicule certain entities. R06 explained, \textit{``There are cases where the noun itself carries hate speech. For example, when I revealed the target like `Democratic Party,' I found nouns [`Democrap'] that people use mockingly. In such cases, when I sensed mockery or discrimination, especially toward race, gender, or nationality, I decided to delete the comment if the expression seemed overly aggressive.''} % [Removed comments with pejorative terms] 명사 자체가 갖는 그런 혐오 표현이 있을 수도 있더라구요. 예를 들어서 Democratic Party을 이렇게 타겟을 이렇게 눌렀을 때 조금 사람들이 조롱하듯이 만든 그런 명사(Democrat)들이 있더라고요. 그런 거에서 조롱의 혹은 차별의 느낌, 그리고 민족, 성별 그리고 국가 이런 거에 대해서 좀 과격한 표현이 있다고 생각되면 삭제를 했습니다. (정다운사자, revealing)

%\noindent\textbf{Aiming for a Comprehensive Understanding of the Overall Content.} 
%Alternatively, some participants considered the overall meaning of the comments, such as keeping those with logical criticism, deleting offensive opinions without valid reasoning, or deleting comments containing fake news. Especially the majority of the participants who encountered less offensive expressions with \offensiveparaphrasing{} (i.e.,~paraphrasing and revealing groups) made moderation decisions by evaluating the broader context of the comments. 
On the other hand, some participants, especially those exposed to milder expressions through \offensiveparaphrasing{}~(i.e.,~paraphrasing and revealing groups), made moderation decisions by \textbf{assessing the detailed context}, such as keeping logical criticism or removing offensive comments without valid reasoning.
P10 described, \textit{``I think I deleted more comments based on the overall content rather than specific words. If the content was fake news or targeted certain regions, countries, or races with hate, I tended to delete those comments first.''} %[Considered overall content] 전반적으로 그냥 전체적인 내용을 보고 삭제를 한 게 더 많았던 것 같아요. 되게 단어보다도 전체적인 내용이 너무 그냥 가짜 뉴스거나 아니면 그냥 무조건 혐오하는 지역이 있거나 국가 인종 이런 것들은 우선 그냥 삭제를 했던 게 맞는 것 같긴 해요. (신바람나는개구리, offensive)

%\noindent\textbf{Selectively Using Revealing Features.} 
Participants in the revealing group mentioned that they \textbf{adopted a selective approach} when using \targetrevealing{} and \offensiverevealing{}. R04 explained that \targetrevealing{} was used where determining whether a comment constituted hate speech depended on the target. Referring to a comment in Fig.~\ref{fig:design}, she noted, \textit{``If the target here had been something like, `It's criminals who bring embarrassing moment or downfall to Korea,' I wouldn't have considered it hate speech. So, in cases like this, I checked the target.''} % [Interaction: [target_revealing] Used when the comments could be hate speech depends on the targets] (역시 대한민국 ※망신※은 ☆여자☆들이 다시킨다 ㅠㅠ // hate_26) 근데 여기에 있는 대상이, 역시 대한민국의 난감한 일 혹은 망신을 범죄자들이 시킨다라고 하면 저는 혐오 표현이 아니라고 생각을 했을 것 같아서 이런 경우에는 대상을 확인했던 것 같습니다. (귀여운토끼, revealing) 
R02 explained he used \offensiverevealing{} only when the flow of altered comments felt awkward, saying \textit{``I only checked the original expression when the sentence didn’t match the previous or following sentences. Otherwise, I could generally understand the comment, so I didn’t often check the original expressions.''} % 읽다가 이게 앞에 그 문장이랑 뒷문장이랑 살짝 매치가 안 될 때 그때 확인을 했고 나머지 때는 대략적으로 좀 파악이 가능해서 표현 완화한 거는 잘 확인을 안 했습니다. (정다운개구리, revealing) 

The reasons for the selective use of revealing features were clustered into two purposes: accurate moderation and mental well-being protection. 
Half of the interviewed participants from the revealing group said they frequently used \targetrevealing{} and \offensiverevealing{}, as they found it challenging to determine whether a comment should be deleted or kept based solely on the anonymized or paraphrased expressions. % (5/10) 아름다운토끼, 정다운개구리, 정다운사자, 신바람나는나비, 활기찬낙타
R01 described \textit{``When I saw only the softened expression, I couldn't really tell whether the comment was something that should be deleted, so I immediately checked the original version.''} % [Interaction: [offensive_revealing] Used in most cases] 후자(완화된 버전을 잘 보지 않고 바로 원본을 확인)인 것 같은 게 원본만 완화된 표현을 봤을 때 이게 진짜 지워야 되는 댓글인지 유지해야 되는 댓글인지 막 저에게 직접적으로 와닿지 않았어서 바로 원본을 확인했습니다. (활기찬낙타, revealing) 
This frequent use of revealing functions was driven by their sense of responsibility as moderators, emphasizing accuracy in their moderation decisions. R02 noted \textit{``... In the end, it was my role to make a judgment, whether this is too hateful, so I checked everything to review the (original) content.''} % [moderation task-driven (accurate)] 사실 저는 다 확인했어요. ... 결국 가치 판단이나 약간 이게 너무 혐오적인가?하는 걸 판단하는 게 제 역할이었으니까 전체적으로 읽으려고 다 확인했습니다. (정다운개구리, revealing) 
% 근데 결국은 이제 댓글 삭제하는 사람은 원본 댓글의 혐오 그 정도로 판단을 해야 되는 거니까 그래서 어느 정도 사실 그 완화를 하는 거는 다 비슷한 결과물이 나오긴 결과물로 완화를 시키는데 엄청 폭력적인 것도 되게 이만큼 완화되고 약간 폭력적인 것도 이만큼 완화되고 하니까 그거를 판단해야 돼서 다 확인했던 원본을 확인해야 됐던 것 같습니다. (아름다운토끼, revealing) 

In contrast to the frequent use of \targetrevealing{} and \offensiverevealing{}, a few participants expressed concerns on the potential emotional impact of confronting offensive expressions. % (2/10) 아름다운곰, 들뜬금붕어
%R05, who used either feature for only fifteen comments, shared \textit{``But somehow, I always hesitated a bit before clicking to reveal the original expression. I kept thinking, `Am I looking at this unnecessarily?' I had that thought quite often ... .''} % 뭔가 근데 이거 약간 그 원본표현을 클릭하기 전에 늘 좀 망설여지는 게 있더라고요. 괜히 보는 거 아닐까? 이런 생각을 되게 많이 했고... (들뜬금붕어, revealing) -- 15/100
R09 explained, \textit{``I felt it was less stressful not to click and reveal everything, so I often just skipped over comments with that kind of (offensive) tone without even looking at the original version. ... I tried to avoid looking at the original as much as possible''}. This participant ended up using the \offensiverevealing{} feature only twice. % 저는 이거를 다 눌러서 보는 것 보다 안눌러서 보는 게 스트레스가 덜하다고 생각이 들어서 ... 이런 말투가 있는 것 들은 (원본을) 보지도 않고 그냥 넘어간 것들이 많아요. ... 요약을 하자면 원본은 되도록이면 많이 안 봤어요. (아름다운곰, revealing) -- 43/100, only 2 offensive
% Interaction: [target_revealing] Tried not to use in order to protect their own emotion // Interaction: [offensive_revealing] Tried not to use in order to protect their own emotion

% \noindent\textbf{Frequent Use of Revealing Features Due to Responsibility as a Moderator.}
% \noindent\textbf{Refraining From Using Revealing Features to Protect Mental Well-Being.} 

The observed strategies illustrate how participants engaged with \system{}’s content modification features to moderate hate speech effectively with less mental burden. Building on these insights, the next sections illustrate how participants described the specific impact of \system{}’s features on moderation accuracy and recall, examining how each feature contributed to or hindered participants' ability to make precise moderation decisions.

\subsubsection{Impact of \system{} on Precise Moderation Decisions}
\label{sec:results:performance:acc-sens-qual}
% We found qualitative evidence suggesting that the anonymizing group had lower average accuracy and \jh{recall} than the control group. 
The anonymizing group often mentioned that they could \textit{``infer the target based on the revealed information, though targets were anonymized''}~(A08).  % 댓글에 그게 타겟이 가려져 있긴 했지만 이제 드러나는 정보들로 위치해 보았을 때, 지역 갈등이나 특정 스포츠에 대한 혐오 아니면 남녀 간의 갈등 이런 걸 좀 조장하는 것 같은 댓글이 많다고 느꼈는데 (산책하는하마, target) 
However, they also reported that \textbf{\targetanony{} hindered accurate moderation} in certain cases, especially when a comment could be interpreted as either an ordinary opinion or hate speech, depending on the target. A04 noted, \textit{``I found it difficult to judge whether the comment should be deleted. Once the anonymized target is revealed, it feels like they might not be genuine hate toward a specific group, but more like fake news.''} % 가려진 경우에 사실 이게 삭제를 해야 되는 건지 말아야 되는 건지에 대한 판단을 하기가 어렵다는 생각이 좀 들었었고요. 그 회색으로 가려진 파트들은 특히 왜냐하면 이게 그걸 공개했을 때 이게 특정 어떤 집단이나 그런 거에 대한 정말 혐오라기보다는 가짜 뉴스 같은 느낌이 들 수도 있는 것 같아서 (산책하는코알라, target) 
Another participant, A06, expanded on the issue, pointing out that certain words could describe factual observations but might still be perceived as hate speech depending on the context: \textit{``Words like `incompetent' are negative, but depending on the target, someone might think it's a fair opinion. For example, saying someone is incompetent could be based on actual data; in that case, deleting the comment doesn't seem right.''} % 무능력이 당연히 부정적인 단어는 맞는데 이제 대상에 따라서 (무능력하다고) 생각할 수도 있을 것 같아요. 예를 들어서 이제 무능력하다는 게 실제로 지표에 근거해가지고 무능력하다 이런 의견이 실제로 있을 수 있잖아요. 그때 이런 걸 삭제하면 이제 좀 그러니까. (수다스러운늑대, target)
This uncertainty around \targetanony{} was also reflected in the lower perceived effectiveness of \targetanony{}, as reported in \S~\ref{sec:results:performance:effectiveness}.
% [target_hiding] Could infer the target based on the context - 산책하는낙타, 산책하는하마 // revealing: 눌러보니까 예측한거 나오더라 (활기찬낙타)
% [target_hiding] Hate speech was distinguishable regardless of visibility of target - 수줍은고양이, 즐거운거위, 산책하는하마 // revealing: 확인하고도 번복하지 않았다 (귀여운토끼, 아름다운곰, 신바람나는나비)
% [target_hiding] Concealing target hindered accurate moderation - 정다운북극곰, 수다스러운늑대, 산책하는코알라, 산책하는낙타, 산책하는하마 (타겟에 따라 결정이 완전히 달라진단 내용이 있음) 

In the paraphrasing group, participants explained they could still sense the nuance of the offensive content. P09 noted, \textit{``Even if an offensive expression is toned down, it still feels like it demeans this group. From the overall context, even if the aggressive wording is reduced, the sentence itself still demeans or disparages a certain group.''} % 삭제 결정 여부는 거의 동일할 것 같아요. 공격적인 거를 완화를 한다고 하더라도 결국에는 이쪽을 폄하하는 게 맞다고 생각이 들거든요. 전체적인 문맥상으로 이게 공격적인 표현을 줄였더라도 이 문장 자체가 한 집단을 폄하하거나 비하하기 때문에... (온화한표범, offensive) 
Some participants mentioned that they \textbf{could infer the original offensive expressions}, saying, \textit{``You know, if it's something like `a person with foggy mind,' you can pretty much guess the kind of insult that's implied''} (P08). % 여러 가지가 있을 것 같은데 정신이 흐릿한 사람 뭐 이런 거면 이제 대충 예상되는 욕이 있잖아요. 비속어만 있는 댓글은 좀 삭제를 하려고 했던 것 같고 ...  (아름다운카멜레온, offensive)
Understanding the overall context and inferring the original offensive expressions behind \offensiveparaphrasing{} could make participants more sensitive to hate speech, requiring careful consideration. Moreover, P02 mentioned that she ``\textit{took action to delete based on imagining the worst-case scenario in many cases.''} % 약간 좀 최악의 경우를 상상하고 삭제 조치를 한 경우가 많았던 것 같습니다. (점잖은기린, offensive) 

% [offensive_paraphrasing] Could sense the nuance of hate speech from the comments] - 온화한표범, 아름다운카멜레온, 신바람나는개구리, 팔팔한하마, 느긋한금붕어, 자유로운금붕어, 당당한호랑이, 점잖은기린 // revealing: 완화되어 있어도 내용 자체가 혐오스러운게 있었음 - 신바람나는팬더, 아름다운곰, 점잖은표범, 정다운사자
% [offensive_paraphrasing] Could infer the original offensive expression - 팔팔한곰, 아름다운카멜레온, 당당한호랑이 // revealing: 그냥 예상되는 것들이 있었다 - 아름다운곰, 들뜬금붕어, 정다운개구리, 점잖은표범
% [offensive_paraphrasing] Made moderation decisions with the assumption that there's a more offensive expression  점잖은기린 // revealing: 좀 더 레디컬한 단어가 있겠다고 짐작하고 눌러보았다 - 신바람나는팬더

Meanwhile, both anonymizing and paraphrasing groups expressed that they wanted to delete some comments that had been kept while reviewing their moderated comment lists during the interview. The comments they referred to often contained pejorative terms, which were anonymized by the \targetanony{} or replaced with neutral targets by the \offensiveparaphrasing{}. For instance, terms such as `Feminazi,' `Ching chong' (a pejorative term for Chinese), and `Nip' (a pejorative term for Japanese) were mentioned by participants. % 꼴펨, 도태남, 왜놈, 짱깨, 더불어적폐당..? 
Participants from the revealing group, who initially encountered modified comments through \targetanony{} and \offensiveparaphrasing{} but had the option to view the original expressions, were \textbf{able to moderate comments containing these pejoratives}, which might explain their higher recall than the other groups. R06 noted, \textit{``For example, when the target is anonymized, technically, it's not considered hate speech. But in cases like calling Japanese people `monkeys,' the noun itself becomes hate speech, and seeing these terms made me decide to delete them.''} % 예를 들어서 이런 것들은 타겟이 없었을 때 사실 혐오 표현은 아니거든요. 그리고 이런 게 진짜 좀 대표적인 예시인 것 같고요. 일본인들을 원숭이라고 이렇게 표현하는 것 자체는 명사가 혐오 표현이 될 수 있으니까 이런 것들은 이 명사를 보고 블록을 해야겠다라고 생각을 했습니다. (정다운사자, revealing)

In summary, participants adopted different strategies depending on the visibility of original expressions, either focusing on explicit expressions or evaluating the broader context. In addition, participants exhibited a bifurcated approach to using the revealing features: some prioritized accuracy in moderation, while others engaged the features selectively to protect their emotional well-being. While \targetanony{} may compromise accurate moderation by limiting information, \offensiveparaphrasing{} encourages a more complex moderation practice, prompting moderators to infer the original expressions and interpret the overall contents carefully. Furthermore, \targetrevealing{} and \offensiverevealing{} facilitated precise moderation by allowing them to verify pejoratives.
% \jh{Participants found it challenging to accurately moderate comments when content was anonymized or paraphrased, as this hindered the recognition of hate speech, especially for comments with pejorative terms, while revealing original expressions facilitated precise moderation. Participants valued the control provided by the revealing features.}
% 

%by sensing the offensive nuance and inferring the original expressions from the paraphrased ones. Furthermore, the revealing group achieved the highest moderation \jh{recall} among groups, benefiting from \offensiveparaphrasing{} at first and effectively moderating comments with pejorative terms by revealing features. 
\section{Discussion}
\label{sec:discussion}
% Forward-looking based on our findings... 

We evaluated \system{}, a text content modification system designed to support hate speech moderation while protecting moderators' mental well-being. \system{} offers four main features: \targetanony{}, which anonymizes the target of hate speech; \offensiveparaphrasing{}, which paraphrases offensive expressions into less harmful language; and \targetrevealing{} and \offensiverevealing{}, which allows users to reveal the target and original expressions with a click. Contrary to our expectations, we did not observe a significant improvement in emotional state or a reduction in fatigue after moderation when comparing the experimental groups with the control group. However, the experimental groups considered modified comments less severe and perceived \system{} to effectively protect their mental well-being. 

Furthermore, the moderation accuracy remained similar despite \system{} modifying the comments by anonymizing the target and paraphrasing offensive expressions into less offensive ones. Notably, the participants who used \offensiveparaphrasing{} showed slightly higher moderation recall. In interviews, our participants described the revealing features of \system{} as a type of \textit{buffer}, providing time for them to prepare to face the offensive original expressions. They also noted that \system{} prevented them from normalizing biased and hateful opinions from the comments by anonymizing targets and paraphrasing offensive expressions. Despite such positive aspects, we did not find clear evidence that \system{} can better safeguard a moderator's overall emotion and fatigue after moderation.  

In this section, we discuss possible explanations for the mixed findings between perceived benefits and actual impacts of \system{} on mental well-being. We then examine in depth how hate speech moderation can still be accurate despite textual modifications, highlighting further considerations for content modification in text content moderation. Additionally, we shed light on possible strategies for protecting moderators' mental well-being for a sustainable working environment.    

% ========================================================================================================
\subsection{Understanding the Discrepancy Between Perceived Benefits and Actual Impact on Mental Well-Being}
\label{sec:discussion:discrepancy}
Our findings revealed a discrepancy between measured outcomes and participants' opinions: while participants perceived comments modified by \system{} as less hateful and positively viewed the effectiveness of \targetanony{} and \offensiveparaphrasing{}, there was no significant difference in the emotion scale or fatigue levels between experimental groups and the control group. Several factors may contribute to this gap between perceived benefits and actual impact on mental well-being.

One potential explanation is that while the paraphrasing and revealing groups evaluated the modified comments as less hateful than the control group, they also took significantly longer to process each comment owing to the uncertainty induced by paraphrasing offensive phrases. Although these features helped lower the perceived severity of each comment, our participants spent a longer time on moderation, resulting in longer cognitive engagement with each potentially hateful comment. This might have contributed to significant changes in emotion and fatigue scales after the experiment. In other words, \system{} reduced the intensity of hate speech, but the longer exposure may have had a negative impact, potentially diminishing the intended emotional protective effect.   

Additionally, the cognitive load associated with inference generation could cause negative emotion and fatigue at the end of the experiment. Our participants indicated that as the target of hate speech was anonymized and the offensive expression was paraphrased, they often inferred and imagined the original expression to understand the full context of a comment and make a moderation decision. This inference process, which requires considerable cognitive effort~\cite{mckoon1992inference}, may have affected participants in the experimental groups, resulting in changes in emotion and fatigue levels similar to those observed in the control group.

Another consideration is the short duration of hate speech exposure in our study. Participants moderated 100 comments, with only 50 containing hate speech, and the control group took an average of 18.07 minutes in total moderation time, with the shortest session lasting only 10.45 minutes. Given this limited exposure, it is possible that the system's protective benefits were not fully realized in such a short session. In contrast, commercial moderators typically work extended hours each day~\cite{roberts2019behind}. A longer moderation session using \system{} could better reveal its protective effects, not only in perceived benefits but also in measurable outcomes for mental well-being.

The results of this study show that text content modification systems such as \system{} have potential, but underlying complexities impact their real-world effectiveness in protecting moderators’ mental well-being. Future studies might examine how these factors (e.g.,~exposure duration, perceived severity, and cognitive load) interact over longer moderation sessions to better understand how text content modification tools can sustainably support moderators' mental well-being.

% ========================================================================================================
\subsection{Text Content Modification in the Context of Content Moderation}
\label{sec:discussion:text}
We found that using \targetanony{} and \offensiveparaphrasing{} maintained comparable moderation accuracy to a control condition, even with anonymized targets and paraphrased offensive expressions. This resilience may stem from \textit{inference generation} processes, where readers actively use their knowledge and context to fill gaps in a text to create logical conclusions~\cite{mckoon1992inference}. Our participants explained that they inferred the original hateful meaning to understand and make moderation decisions, mentally reconstructing the likely intent of paraphrased or concealed expressions. This ability to \textit{fill in} the modified elements of hate speech, even when information was limited, allowed participants to maintain accuracy comparable to that of the control group. 
% As noted by P28, \textit{``you can pretty much guess the kind of insult that’s implied''}.

Another unexpected finding was that participants in the paraphrasing and revealing groups, who moderated comments using \offensiveparaphrasing{}, exhibited slightly higher recall than the control group. This may also be attributed to the inference generation process, as participants took more time to reconstruct the possible offensive intent behind paraphrased comments, resulting in more careful judgments. This aligns with the dual process theory of human decision making~\cite{norman1986attention}, which is also known as the concept of \textit{thinking fast and slow}---system 2's slow thinking involves a deeper and more thoughtful evaluation of statements and their implications, whereas system 1's fast thinking lacks such deliberate reasoning~\cite{kahneman2011thinking,choi2024foodcensor}.

Anonymizing targets or paraphrasing offensive phrases fosters a reflective approach to moderation, as it introduces uncertainty that requires moderators to actively infer concealed targets or offensive meanings. We posit that this uncertainty in user interactions facilitates mindful interaction, similar to traditional `interaction lockout' or `friction' designs, which restrict user interactions for safety or prevent human errors~\cite{cox2016design}. Our moderation features can introduce a lightweight interaction lockout that slows cognitive processing and encourages deliberate thought. 

While uncertainty or ambiguity in HCI literature has historically been explored to inspire design and enrich hedonic interactions~\cite{gaver2003ambiguity}, its role is expanding. Traditionally, designers introduced elements of ambiguity or uncertainty (e.g.,~information, context, and relationship) to foster curiosity, enhance engagement, and encourage self-reflection, particularly in creative and playful systems. Uncertainty also plays an important role in pragmatic applications that involve data contextualization and sense-making~\cite{kim2024navigating}. In content moderation contexts, such elements of uncertainty not only create a psychological buffer for content moderation but also possibly improve the performance of content moderation (e.g.,~higher recall in hate speech moderation). 

While our study underscores the potential of text content modification for hate speech moderation, additional directions remain for exploration. One possible consideration is the question of authorship in modified user-generated content. Given that \system{} presents content in its modified form for evaluation, users might argue that moderation decisions do not accurately reflect the original intent or expression behind what they had written. Authorship of LLM-based paraphrased content has been actively discussed in AI and Human-AI Interaction fields~\cite{tripto2023ship,yuan2022wordcraft}, where authorship typically depends on two factors: \textit{content}, representing the subject matter of thematic focus, and \textit{style}, the distinctive manner of expression~\cite{sari2018topic}. Recent research has argued that LLM-based paraphrasing retains the core \textit{content} while altering the style~\cite{tripto2023ship}. Traditional views on authorship often emphasize the author's unique ideas, concepts, or thoughts, aligning more closely with \textit{content}~\cite{samuels1988idea,foucault2003author}. Given that \system{} preserves core \textit{content} (i.e.,~hateful intent) while modifying \textit{style} to reduce offensive language, moderation decisions based on the modified content could reasonably translate to judgments on the original version. However, further investigation is needed to understand how users perceive having their content assessed in a modified form and how they might react to decisions based on these modifications.

% In line with this, another future direction could be enhancing transparency in moderation decision notifications. Currently, users often perceive these notifications as opaque~\cite{myers2018censored}, and decisions based on modified content could further exacerbate this issue. One possible design solution could involve moderators drafting notifications using anonymized and paraphrased content, where the system could automatically revert these expressions to the original wording to offer clearer reasoning for users.

Additionally, when users request a reconsideration of decisions they find incorrect or unfair, the appeals process generally allows them to explain in free text why they believe the decision was incorrect~\cite{tspa}. However, text content modification through \system{} introduces an information asymmetry: moderators view the modified text, while appealers present arguments based on their original content. Further research exploring the impact of this information disparity could be valuable in guiding the development of a fair and transparent appeal process for situations where text content modification techniques are used for moderation.

% Unlike previous approaches that reduce visual stimuli through techniques such as grayscaling, blurring, or cartooning images and videos~\cite{}, our approach directly limits key information within the text by changing hate speech-related signals. For example, anonymizing target words is akin to covering violent areas in images with a ``black box,'' while paraphrasing offensive expressions resembles substituting a problematic object with a similar, less provocative one. This raises an intriguing question: How did moderators still achieve comparable accuracy despite such concealment of key hate speech indicators?

\subsection{Toward a Sustainable Career in Content Moderation}
\label{sec:discussion:sustainable}
Our qualitative findings highlight the potential for text content modification in hate speech moderation to support moderators' mental well-being. Participants discussed how \system{} protected their personal viewpoints from being shaped by biased and hateful opinions from the comments. Unlike image and video moderation, where explicit visual stimuli (e.g.,~violent or sexual content) can trigger immediate, sensory-based reactions~\cite{uhrig2016emotion}, text-based hate speech presents a distinct set of concerns via repeatedly exposing moderators to biased or extreme opinions. Moderating this type of content requires individuals to engage in a complex cognitive process to interpret and reconstruct the underlying message, as discussed in \S\ref{sec:discussion:text}, which may have long-term effects on their personal perspectives. %These unique characteristics of text-based hate speech underscore the importance of targeted support in preserving moderators' mental well-being, particularly within text domains. 

% HateBuffer might protect moderators from sensitization & desensitization in a long-term. 
Repeated and prolonged exposure to such violent, negatively stimulating content places moderators at risk of longer-term psychological damage, including emotional desensitization~---~characterized by a gradual numbing of emotional reactions and reduced empathy toward real-world situations~\cite{linz1984effects,krahe2011desensitization}~---~or, conversely, emotional sensitization, wherein repeated exposure heightens their responsiveness to stressful stimuli~\cite{blumstein2016habituation}. Recent work has documented instances where experienced moderators exhibited adverse reactions even to traditionally positive emotional stimuli intended to alleviate stress, underscoring the complexities of sensitization and its challenges for effective emotional support~\cite{cook2022awe}. Although our study did not directly measure sensitization or desensitization, participants indicated that content modification, anonymizing target expressions, and/or paraphrasing offensive expressions could potentially reduce their mental burden. Such support may help moderators sustain their roles over a longer period, creating a more stable working environment and reducing turnover rates commonly observed in moderation work~\cite{guardian2024the}.
% SB: desensitization/sensitization이 궁극적으로 사람에게 주는 문제 (e.g., 삶에서 감정에 무뎌짐) 같은걸 언급해야하나? 언급하지 않아도 자명하긴한데... 

The revealing features of our study allowed participants to selectively view the targets and original offensive expressions, providing a sense of control. Facebook’s Global Resiliency Team has noted that \textit{``shielding moderators from harm begins with giving them more control over what they see and how they see it''}~\cite{mark2019facebook}. Given that a lower sense of control in high-stress workplaces can increase stress levels~\cite{akbari2017job}, tools such as \system{} could enhance moderators' agency over their exposure to potentially hateful text-based content. In addition, our participants found revealing features helpful in providing them time to mentally prepare before encountering potentially distressing language, consistent with findings of previous work~\cite{kim2024respect}. Considering the repetitive and emotionally demanding nature of moderation~\cite{roberts2019behind,steiger2021psychological}, this sense of control supports self-efficacy and promotes healthier engagement with the work.

With that said, implementing content modification tools such as \system{} in real-world moderation settings involves navigating a tension between performance and moderator well-being. Our findings show that reviewing modified content, designed to reduce the offensiveness of hate speech, led to longer task completion time while maintaining quantitative fatigue levels. These outcomes are consistent with observations with commercial moderators, who often prefer reviewing a smaller number of severe cases over a high volume of mildly offensive content, citing the increased cognitive load and time demands of the latter~\cite{strongylou2024perceptions}. In such settings, especially where strict performance quotas are in place~\cite{narayanan2024ai,wohn2019volunteer,caimoderation2021}, the need to interpret softened expressions more carefully may contribute to slower moderation. However, given that commercial platforms typically provide far more detailed moderation guidelines than those used in our study~\cite{narayanan2024ai}, it remains unclear how such content modifications would affect moderators' speed and experience in practice. Future work should examine how tool use, time constraints, and policy design interact to shape moderation outcomes and moderator well-being in real-world environments.

In addition, prior research in content moderation has emphasized the importance of providing appropriate rest, which can benefit moderators' mental well-being~\cite{cook2022awe,steiger2021psychological,faucett2007rest}; structured breaks are essential in maintaining workplace well-being and reducing fatigue~\cite{blasche2017effects}. While a standard recommendation for break duration is known as a 7.5-minute break after 50 minutes of content review~\cite{boucsein1997design}, break scheduling could be further adjusted based on the frequency of exposure to original, unmodified content, especially when content modification techniques are used in text moderation. %For instance, moderators could work continuously for longer periods when primarily reviewing modified content, while shorter sessions with more frequent breaks may be more suitable when reviewing original offensive content. 
Future research should explore the optimal balance of modified and original content exposure and break frequency and duration to effectively support moderators' mental well-being.

\subsection{Limitations and Future Work}
\label{sec:discussion:limitation}
In this paper, we present insights from a mixed-methods approach, combining quantitative analysis with in-depth qualitative insights, which allows us to explore the potential for text content modification to support moderators' mental well-being in hate speech moderation. In addition, our user study was conducted using a simulated moderation task with pre-curated data. This controlled setup enabled us to protect participants from encountering unexpected harmful content, but it may not fully capture the range and complexity of real-world hate speech moderation scenarios. Implementing a full pipeline that includes AI-based detection of targets and offensive expression through techniques such as entity recognition~\cite{li2020survey} and sentiment analysis~\cite{wankhade2022survey}, followed by paraphrasing via LLMs could facilitate field studies with real-world data. Moreover, evaluating the actual moderation accuracy with \system{} through large-scale studies in real-world settings could provide deeper insights into the practical applicability of content modification. Such studies could offer more comprehensive insights into \system{}'s effectiveness within content moderation's dynamic, unpredictable nature. 

Additionally, our simulated moderation experiment utilized the K-HATERS dataset~\cite{park2023k}, which consists of news article comments that may lack full context. Although we made efforts to select comments that seemed understandable independently of their original news articles, there remains a possibility that participants might not fully grasp certain context-dependent meanings. However, context dependency is not unique to news comments; moderation of content from other platforms, such as X's threads, Reddit's discussions, or YouTube's comments, may similarly face challenges if moderators review comments without the original root content. Therefore, investigating how moderation performance and moderators' experiences with \system{} vary across diverse content types and platforms, including different degrees of contextual availability, would be a valuable direction for future research.

Furthermore, since the K-HATERS dataset includes topics specifically relevant to Korean society, such as gender issues and internal regional discrimination, our findings may have limited generalizability across different cultural contexts. Hate speech varies significantly across cultures, reflecting unique social, political, and historical factors~\cite{gandhi2024hate}. For example, prevalent topics of hate speech in other cultures, such as gun control and immigration, were not present in our dataset. Even for the same topic, the targeted race or event may differ~\cite{castano2021internet}. To gain a broader understanding of text content modification systems' impact on hate speech moderation, it would be valuable to investigate their effects across various cultural contexts, expanding our observations to account for global dynamic social issues.
Moreover, while \system{}'s \offensiveparaphrasing{} was designed to paraphrase hate speech effectively, a key challenge remains in how accurately LLMs can interpret and respond to cultural nuances. As the types, targets, and subtleties of hate speech vary widely across cultures, incorporating a culturally adaptive or localized model~\cite{narula2024comprehensive} could enhance \system{}'s ability to moderate contextually relevant hate speech and increase its applicability across diverse cultural backgrounds.

\section{Conclusion}
\label{sec:conclusion}

We designed \system{}, a text-based content modification system for hate speech moderation, to safeguard moderators' mental well-being while preserving moderation performance. We conducted a user study with 80 participants who were assigned the role of moderators to perform simulated hate speech moderation from a fictional news platform and observed qualitative insights through semi-structured interviews.
In contrast to our expectation, we could not observe any improvement in emotion and fatigue after hate speech moderation with \system{} compared with the control group. However, the perceived hate severity of comments was significantly lower when \system{} is used, and participants recognized \system{} as an effective \textit{buffer} for emotional contagion and normalization of biased opinions from hate speech. Notably, \system{} did not hinder the moderation accuracy, even enabling slightly higher recall. 
Building on these findings, we explored possible explanations for the discrepancy between perceived benefits and the actual impact of \system{} on mental well-being. We highlighted the potential of the content modification technique in the text as a content moderation tool and mental well-being protection tool, fostering a more sustainable working environment for content moderators.

\begin{acks}
This work was supported by the Institute of Information \& communications Technology Planning \& Evaluation (IITP) grant funded by the Korea government (MSIT) (No.RS-2025-02263169, Detection and Prediction of Emerging and Undiscovered Voice Phishing), funded by the Institute of Information \& communications Technology Planning \& Evaluation (IITP) grant funded by the Korean government (MSIT) (RS-2022-II220064), and funded by the National Research Foundation of Korea (NRF) grant funded by the Korea government (MSIT) (RS-2024-00348993).
\end{acks}

\newpage
\bibliographystyle{ACM-Reference-Format}
\bibliography{main}

\newpage
\onecolumn
\appendix
% \section{Supplementary Material}

\section{Full Prompt for Paraphrasing Comments}
\label{sec:supp:prompt}

\subsection{Paraphrasing Hate Comments}
You are a text content moderator for Korean news comments. Your task is to paraphrase the original comment into a euphemistic comment while maintaining the original meaning of the comment. For your information, Euphemism is an innocuous word or expression used in place of one that is deemed offensive or suggests something unpleasant. From a given Korean comment on the News, output 10 distinct Euphemistic paraphrased comments by paraphrasing only words between `※'. Do not change the original meaning and separate each output with `\$'.

\subsection{Paraphrasing Normal Comments}
You are a text content moderator for Korean news comments. Your task is to paraphrase a comment. From a given Korean comment on the News, output 10 distinct paraphrased comments by paraphrasing only words between `※'. Do not change the original meaning and separate each output with `\$'.

\section{Shapiro-Wilk Test for SPANE\_B Scores: Pre- and Post-Experiment}
\label{sec:supp:spaneb}

% \usepackage{multirow}
\subsection{Pre-experiment}
\label{sec:supp:spaneb_pre}
\begin{center}
% \begin{table}[h]
    % \centering
    \captionof{table}{Shapiro-Wilk Test Results of Pre-experiment scores for SPANE\_B}
    \label{tab:spane_pre}
\begin{tabular}{ccccccc}
\hline
\multirow{2}{*}{\textbf{Group}} & \multirow{2}{*}{\textbf{Mean}} & \multirow{2}{*}{\textbf{Std}} & \multirow{2}{*}{\textbf{Min}} & \multirow{2}{*}{\textbf{Max}} & \multicolumn{2}{c}{\textbf{Shapiro-Wilk}} \\ \cline{6-7} 
                       &                       &                      &                      &                      & \textbf{W}             & \textbf{p-value}          \\ \hline
Control                & 12.00                 & 5.38                 & 3.00                 & 20.00                & 0.91          & .625             \\
Anonymizing            & 10.65                 & 4.86                 & 1.00                 & 19.00                & 0.97          & .659             \\
Paraphrasing           & 10.35                 & 5.03                 & 3.00                 & 19.00                & 0.94          & .256             \\
Revealing              & 10.35                 & 7.84                 & -4.00                & 24.00                & 0.96          & .587             \\ \hline
\end{tabular}
\end{center}
% \end{table}

% \usepackage{multirow}
\subsection{Post-experiment}
\label{sec:supp:spaneb_post}
% \begin{table}[h]
\begin{center}
    % \centering
    \captionof{table}{Shapiro-Wilk Test Results of Post-experiment scores for SPANE\_B}
    \label{tab:supp:spane_post}
\begin{tabular}{ccccccc}
\hline
\multirow{2}{*}{\textbf{Group}} & \multirow{2}{*}{\textbf{Mean}} & \multirow{2}{*}{\textbf{Std}} & \multirow{2}{*}{\textbf{Min}} & \multirow{2}{*}{\textbf{Max}} & \multicolumn{2}{c}{\textbf{Shapiro-Wilk}} \\ \cline{6-7} 
                       &                       &                      &                      &                      & \textbf{W}             & \textbf{p-value}          \\ \hline
Control                & 1.95                  & 7.17                 & -11.00               & 13.00                & 0.96          & .478             \\
Anonymizing            & 2.70                  & 7.19                 & -13.00               & 14.00                & 0.95          & .420             \\
Paraphrasing           & 0.20                  & 7.12                 & -17.00               & 14.00                & 0.92          & .096             \\
Revealing              & -0.30                 & 6.20                 & -10.00               & 13.00                & 0.92          & .109             \\ \hline
\end{tabular}
\end{center}
% \end{table}

\section{Shapiro-Wilk Test for \fatigue{} Scores: Pre- and Post-Experiment}
\label{sec:supp:fatigue}

% \usepackage{multirow}
\subsection{Pre-experiment}
\label{sec:supp:fatigue_pre}
% \begin{table}[h]
\begin{center}
    % \centering
    \captionof{table}{Shapiro-Wilk Test Results of pre-experiment scores for \fatigue{}}
\begin{tabular}{ccccccc}
\hline
\multirow{2}{*}{\textbf{Group}} & \multirow{2}{*}{\textbf{Mean}} & \multirow{2}{*}{\textbf{Std}} & \multirow{2}{*}{\textbf{Min}} & \multirow{2}{*}{\textbf{Max}} & \multicolumn{2}{c}{\textbf{Shapiro-Wilk}} \\ \cline{6-7} 
                       &                       &                      &                      &                      & \textbf{W}             & \textbf{p-value}          \\ \hline
Control                & 2.60                  & 8.12                 & 17.00                & -11.00               & 0.96          & .478             \\
Anonymizing            & 2.25                  & 6.67                 & 13.00                & -7.00                & 0.91          & .072             \\
Paraphrasing           & 2.80                  & 8.32                 & 23.00                & -9.00                & 0.96          & .482             \\
Revealing              & 3.75                  & 7.91                 & 18.00                & -12.00               & 0.97          & .840             \\ \hline
\end{tabular}
\end{center}
% \end{table}

% \usepackage{multirow}
\subsection{Post-experiment}
\label{sec:supp:fatigue_post}
% \begin{table}[h]
\begin{center}
    % \centering
    \captionof{table}{Shapiro-Wilk Test Results of post-experiment scores for \fatigue{}}
\begin{tabular}{ccccccc}
\hline
\multirow{2}{*}{\textbf{Group}} & \multirow{2}{*}{\textbf{Mean}} & \multirow{2}{*}{\textbf{Std}} & \multirow{2}{*}{\textbf{Min}} & \multirow{2}{*}{\textbf{Max}} & \multicolumn{2}{c}{\textbf{Shapiro-Wilk}} \\ \cline{6-7} 
                       &                       &                      &                      &                      & \textbf{W}             & \textbf{p-value}          \\ \hline
Control                & 9.05                  & 9.02                 & 31.00                & -6.00                & 0.96          & .540             \\
Anonymizing            & 6.20                  & 7.18                 & 23.00                & -7.00                & 0.97          & .777             \\
Paraphrasing           & 8.35                  & 9.32                 & 28.00                & -10.00               & 0.99          & .988             \\
Revealing              & 10.50                 & 8.13                 & 23.00                & -8.00                & 0.96          & .607             \\ \hline
\end{tabular}
\end{center}
% \end{table}

\section{Wilcoxon Signed-Rank Test Results comparing Pre- and Post-experiment for SPANE and \fatigue{}}
\label{sec:supp:wilcoxon}

{\setlength{\parskip}{0pt}}
\subsection{Pre- and Post-experiment for SPANE}
\label{sec:supp:spaneb_wilcoxon}
% \begin{table}[h]
\begin{center}
    % \centering
    \captionof{table}{Wilcoxon Signed-Rank Test Results comparing pre- and post-experiment scores for SPANE questionnaire items.}
\begin{tabular}{lll}
\hline
\textbf{Comparison} & \textbf{Statistics} & \textbf{p-value}    \\ \hline
Control             & 0.00                & \textless{}.001*** \\
Anonymizing         & 8.50                & \textless{}.001*** \\
Paraphrasing        & 0.00                & \textless{}.001*** \\
Revealing           & 3.00                & \textless{}.001*** \\ \hline
\end{tabular}
\end{center}
% \end{table}

\subsection{Pre- and Post-experiment for \fatigue{}}
\label{sec:supp:fatigue_wilcoxon}
% \begin{table}[h]
\begin{center}
    % \centering
    \captionof{table}{Wilcoxon Signed-Rank Test Results comparing pre- and post-experiment scores for \fatigue{} questionnaire items.}
\begin{tabular}{lll}
\hline
\textbf{Comparison} & \textbf{Statistics} & \textbf{p-value} \\ \hline
Control             & 22.50               & .002**          \\
Anonymizing         & 28.00               & .012*            \\
Paraphrasing        & 17.50               & .002**          \\
Revealing           & 21.50               & .002**          \\ \hline
\end{tabular}
\end{center}
% \end{table}

\section{Survey Questionnaire}
\label{sec:supp:survey}
% post 기준으로 작성하고, pre에서는 ... 만 물어봤다. 라고 적기
% https://docs.google.com/document/d/1tmzyJ8G_pY-4AY_UC7dKpOYuL3F7kEdYhp1-byw4WUQ/edit?usp=sharing
This survey contains the following two sections.
\begin{enumerate}[topsep=1pt,noitemsep]
    \item[(1)] Korean Scale of Positive and Negative Experience (pre- and post-survey)
    \item[(2)] Multidimensional Fatigue Symptom Inventory (pre- and post-survey)
    \item[(3)] System Evaluation Questionnaire (post-survey)
\end{enumerate}
\noindent{Please answer all questions sincerely.} \\

\noindent\textbf{Korean Scale of Positive and Negative Experience}

This scale consists of a number of words that describe different feelings and emotions. Read each item and indicate to what extent you feel at this moment before you have started the experiment (at this moment after you have finished the experiment). [1: Not at All, 2: A Little, 3: Moderately, 4: Quite a Bit, 5: Extremely]

\begin{table}[H]
\begin{tabularx}{\textwidth}{X X}
\begin{tabular}{@{}l@{}} % Left side
(1) Positive   \\ 
(2) Negative   \\ 
(3) Good       \\ 
(4) Bad        \\ 
(5) Pleasant   \\ 
(6) Unpleasant 
\end{tabular} 
& % Right side
\begin{tabular}{@{}l@{}} 
(7) Happy      \\ 
(8) Sad        \\ 
(9) Afraid     \\ 
(10) Joyful    \\ 
(11) Angry     \\ 
(12) Contented 
\end{tabular}
\end{tabularx}
\end{table}

% \begin{enumerate}[topsep=1pt,noitemsep]
%     \item[(1)] Positive
%     \item[(2)] Negative
%     \item[(3)] Good
%     \item[(4)] Bad
%     \item[(5)] Pleasant
%     \item[(6)] Unpleasant
%     \item[(7)] Happy
%     \item[(8)] Sad
%     \item[(9)] Afraid
%     \item[(10)] Joyful
%     \item[(11)] Angry
%     \item[(12)] Contented 
% \end{enumerate}

\noindent\textbf{Multidimensional Fatigue Symptom Inventory}

Below is a list of statements that describe how people sometimes feel. Please read each item carefully, then mark the one number which best describes how true each statement is for you at this moment before you have started the experiment (at this moment after you have finished the experiment). [1: Not at All, 2: A Little, 3: Moderately, 4: Quite a Bit, 5: Extremely]

\begin{table}[H]
\begin{tabularx}{\textwidth}{X X}
\begin{tabular}{@{}l@{}} % Left side
(1) I have trouble remembering things   \\ 
(2) I feel upset   \\ 
(3) I feel cheerful       \\ 
(4) I feel lively        \\ 
(5) I feel nervous   \\ 
(6) I feel relaxed     \\  
(7) I am confused     \\
(8) I feel sad     \\
(9) I have trouble paying attention     \\
\end{tabular} 
& % Right side
\begin{tabular}{@{}l@{}} 
(10) I am unable to concentrate      \\ 
(11) I feel depressed        \\ 
(12) I feel refreshed     \\ 
(13) I feel tense    \\ 
(14) I feel energetic     \\ 
(15) I make more mistakes than usual     \\ 
(16) I am forgetful     \\
(17) I feel calm     \\
(18) I am distressed     \\
\end{tabular}
\end{tabularx}
\end{table}

\noindent\textbf{System Evaluation Questionnaire}

Based on your experience using the system during the experiment, please indicate how much you agree or disagree with the following statements, and provide answers to the open-ended questions about your overall comment moderation experience. [1: Not at All, 2: A Little, 3: Moderately, 4: Quite a Bit, 5: Extremely]

\begin{itemize} [topsep=1pt,noitemsep]
    \item \noindent\textit{Control group}
        \begin{enumerate}[topsep=1pt,noitemsep]
            \item[(1)] I had no trouble understanding the meaning of the comments.
            \item[(2)] Please describe the overall process you went through when reading the comments and making deletion decisions. How did you feel when reading the comments? What factors did you consider when deciding whether to delete them? [\textit{open-ended}]
        \end{enumerate}
    \item \noindent\textit{Anonymizing group}
        \begin{enumerate}[topsep=1pt,noitemsep]
            \item[(1)] The fact that the targets were hidden was helpful in performing the comment moderation task. 
            \item[(2)] The fact that the targets were hidden was helpful in maintaining my mental well-being. [\textit{open-ended}]
            \item[(3)] Please describe the overall process you went through when reading the comments with hidden targets and making deletion decisions. How did you feel when reading these comments? What factors did you consider when deciding whether to delete them? [\textit{open-ended}]
        \end{enumerate}
    \item \noindent\textit{Paraphrasing group}
        \begin{enumerate}[topsep=1pt,noitemsep]
            \item[(1)] The following questions pertain to the mitigated offensive expressions in the comments.
            \begin{enumerate}[topsep=1pt,noitemsep]
                \item[(a)] The fact that the expressions were mitigated was helpful in performing the comment moderation task.
                \item[(b)] The fact that the expressions were mitigated was helpful in maintaining my mental well-being.
            \end{enumerate}
            \item[(2)] Please describe the overall process you went through when reading the mitigated comments and making deletion decisions. How did you feel when reading these comments? What factors did you consider when deciding whether to delete them? [\textit{open-ended}]
        \end{enumerate}
    \item \noindent\textit{Revealing group}
        \begin{enumerate}[topsep=1pt,noitemsep]
            \item[(1)] The following questions pertain to the target hiding feature in comments.
            \begin{enumerate}[topsep=1pt,noitemsep]
                \item[(a)] The “view original target” feature was helpful in performing the comment moderation task.
                \item[(b)] The “view original target” feature was helpful in maintaining my mental well-being.
            \end{enumerate}
            \item[(2)] The following questions pertain to the mitigated offensive expressions in comments.
            \begin{enumerate}[topsep=1pt,noitemsep]
                \item[(a)] The “view original expression” feature was helpful in performing the comment moderation task.
                \item[(b)] The “view original expression” feature was helpful in maintaining my mental well-being.
            \end{enumerate}
            \item[(3)] Please describe the overall process you went through when reading the modified comments and making deletion decisions. How did you feel when reading these comments? What factors did you consider when deciding whether to delete them? [\textit{open-ended}]
        \end{enumerate}
        \end{itemize}

\section{Interview Protocol}
\label{sec:supp:interview}
% https://docs.google.com/document/d/11M73crl8F8chQvrqmEm-v9ZggDRhlfS0eyMO3sp9q-s/edit?tab=t.0
We will now begin the interview about the experiment you participated in today.
If there are any questions you don’t wish to answer during the interview, feel free to refuse to answer.

\noindent\textbf{Warm-up and Overall Experience}

First, I will ask some questions about your overall experience with comment moderation.
\begin{enumerate}[topsep=1pt,noitemsep]
    \item[(1)] What was the general intensity of hate speech you felt in the comments you reviewed during the experiment?
    \begin{enumerate}[topsep=1pt,noitemsep]
        \item[(a)] (If negative) You mentioned that the atmosphere of the comments was {what the participant described}. How did moderating such comments affect your emotions or mental well-being?
        \item[(b)] If you had to moderate these comments daily as a comment moderator, how would it impact your emotions or mental well-being?
    \end{enumerate}
    \item[(2)] (Questions asked again about any parts the participant did not elaborate on in the survey)
    \begin{enumerate}[topsep=1pt,noitemsep]
        \item[(a)] How did you feel when reading the comments?
        \item[(b)] What are your criteria for determining hate speech?
        \item[(c)] Were there any difficult instances in deciding whether to delete a comment? Please explain the situation and the reason.
    \end{enumerate}
\end{enumerate}

\noindent\textbf{About the System}

I will now ask some questions regarding the system you used today and comment moderation.
\begin{itemize} [topsep=1pt,noitemsep]
    \item \textit{Control group} 
    
    The system you used today annotated words that could be considered targets of hate speech and offensive expressions in the comments.
        \begin{enumerate}[topsep=1pt,noitemsep]
            \item[(1)] Did the fact that targets were annotated help you determine hate speech? Why or why not?
            \item[(2)] Did the fact that offensive expressions were annotated help you moderate hate speech? Why or why not?
            \item[(3)] Did the fact that targets were annotated affect your mental well-being? If it did, was the effect positive or negative? Please explain the reason.
            \item[(4)] Did the fact that offensive expressions were annotated affect your mental well-being? If it did, was the effect positive or negative? Please explain the reason.
        \end{enumerate}
    \item \textit{Anonymizing group} 
    
    The system you used today concealed words that could be considered targets of hate speech and underlined offensive expressions in the comments.
        \begin{enumerate}[topsep=1pt,noitemsep]
            \item[(1)] Would you make the same moderation decisions even if the targets were not concealed? Why or why not?
            \item[(2)] Do you think reviewing and moderating comments with concealed targets affects your mental well-being differently compared to seeing the original comments? Please explain the reason.
            \item[(3)] If there is a feature to reveal the concealed targets, how would you use it in the decision-making process for comment moderation? If you think you wouldn’t use it, please explain why.
            \item[(4)] Did the fact that offensive expressions were annotated help you moderate hate speech? Why or why not?
            \item[(5)] Did the fact that offensive expressions were annotated affect your mental well-being? If it did, was the effect positive or negative? Please explain the reason.
        \end{enumerate}
    \item \textit{Paraphrasing group} 
    
    The system you used today paraphrased offensive expressions in less offensive and annotated words that could be considered targets of hate speech.
        \begin{enumerate}[topsep=1pt,noitemsep]
            \item[(1)] Would you make the same moderation decisions if you saw the original, unmodified comments? Why or why not?
            \item[(2)] Have you changed your moderation decision after reading another version of the expression using the refresh feature? Please describe the situation and explain why you changed your decision.
            \item[(3)] Would reviewing and moderating comments with the original offensive expressions affect your mental well-being differently than their paraphrased versions? Please explain the reason as well.
            \item[(4)] Suppose there is a feature that allows you to see the original versions of the paraphrased offensive expressions. How would you use it in the decision-making process for comment moderation? If you feel you wouldn’t use it, please explain why.
            \item[(5)] Did the fact that targets were annotated help you determine hate speech? Why or why not?
            \item[(6)] Did the fact that targets were annotated affect your mental well-being? If it did, was the effect positive or negative? Please explain the reason.
        \end{enumerate}
    \item \textit{Revealing group} 
    
    The system you used today concealed words that could be considered targets of hate speech and paraphrased offensive expressions into less offensive ones. It also allowed you to view the original comments or different versions of paraphrased expressions when needed.
        \begin{enumerate}[topsep=1pt,noitemsep]
            \item[(1)] First, I will ask questions related to your moderation decision.
            \begin{enumerate}[topsep=1pt,noitemsep]
                \item[(a)] Was there a situation where you changed your moderation decision after checking the concealed target? Please explain the situation and why you changed your decision.
                \item[(b)] Was there a situation where you changed your decision to delete a comment after checking the original version of the mitigated offensive expression? Please explain the situation and why you changed your decision.
                \item[(c)] Was there a situation where you used the refresh feature to switch to a different paraphrased version of a comment and subsequently changed your decision to moderate it? Please describe the situation and explain why you changed your decision.
                \item[(d)] Among the features—target hiding, offensive expression paraphrasing, refresh, and view original—which was the most helpful in comment moderation, and why?
            \end{enumerate}
            \item[(2)] Next, I will ask questions related to your mental well-being.
            \begin{enumerate}[topsep=1pt,noitemsep]
                \item[(a)] Did checking the concealed targets affect your mental well-being? If so, how did it change, and why?
                \item[(b)] Did checking the original version of the mitigated offensive expressions affect your mental well-being? If so, how did it change, and why?
                \item[(c)] Did using the refresh feature to change the paraphrased versions of the comments affect your mental well-being? If so, how did it change, and why?
                \item[(d)] Among the features—target hiding, offensive expression paraphrasing, refresh, and view original—which was the most helpful in protecting your mental well-being, and why?
            \end{enumerate}
        \end{enumerate}
\end{itemize}

\noindent\textbf{Wrap-up}

We’ve covered almost all the questions about today’s experiment, but I’d like to ask you a simple question before we finish.

\begin{itemize}
    \item (\textit{Control group}) It is known that moderators are often exposed to content that negatively impacts mental health, potentially leading to vicarious trauma or PTSD. Considering the negative effects of moderating comments, are there any features you think should be added to a moderation system?
    \item (\textit{Target, Offensive, Revealing group}) It is known that moderators are often exposed to content that negatively impacts mental health, potentially leading to vicarious trauma or PTSD. The system you used today was designed to help protect moderators. Are there any features you would like to see improved? Or are there any new features you would propose to better preserve the moderator’s mental well-being?
\end{itemize}

Thank you for answering all of our questions thoroughly.
We’ve asked everything we intended to cover. Before we conclude, is there anything you’d like to share with us or any responses you couldn’t fully express during the interview?

Thank you.
This concludes today’s experiment.
We will follow up with additional documents via email for your compensation.

Thank you again.

\end{document}